\documentclass[a4paper,fleqn,usenatbib]{mnras}

\usepackage{newtxtext,newtxmath}
\usepackage[T1]{fontenc}
\usepackage{ae,aecompl}
\usepackage{graphicx}	
\usepackage{enumitem}
\usepackage{rotating}

\newcommand{\te}{\emph{TESS}}
\newcommand{\gc}{$\gamma$\,Cas}
\newcommand{\kms}{km\,s$^{-1}$}

\title[Binarity of \gc\ stars]{Velocity monitoring of \gc\ stars reveals their binarity status\thanks{Based on spectra obtained with TIGRE-HEROS, VLT-UVES, and CAHA-CARMENES}}

\author[Y. Naz\'e et al.]{Ya\"el~Naz\'e$^1$\thanks{F.R.S.-FNRS Senior Research Associate, email: ynaze@uliege.be}, Gregor Rauw$^1$, Stefan Czesla$^2$, Myron A. Smith$^3$, and Jan Robrade$^2$ 
\\
$^1$ Groupe d'Astrophysique des Hautes Energies, STAR, Universit\'e de Li\`ege, Quartier Agora (B5c, Institut d'Astrophysique et de G\'eophysique), \\
All\'ee du 6 Ao\^ut 19c, B-4000 Sart Tilman, Li\`ege, Belgium\\
$^2$ Hamburger Sternwarte, Universit\"at Hamburg, Gojenbergsweg 112, 21029 Hamburg, Germany\\
$^3$ NSF OIR Lab, 950 N Cherry Ave, Tucson, AZ 85721, USA
}

\pubyear{2021}

\begin{document}
\label{firstpage}
\pagerange{\pageref{firstpage}--\pageref{lastpage}}
\maketitle

\begin{abstract}
  The binary status of \gc\ stars has been discussed while theoretically examining the origin of their peculiar X-ray emission. However, except in two cases, no systematic radial velocity monitoring of these stars had been undertaken yet to clarify their status. We now fill this gap using TIGRE, CARMENES, and UVES high-resolution spectroscopy. Velocities were determined for 16 stars, revealing shifts and/or changes in line profiles. The orbit of six new binaries could be determined: the long periods (80--120\,d) and small velocity amplitudes (5--7\,\kms) suggest low mass companions (0.6--1\,M$_{\odot}$). The properties of the known \gc\ binaries appear similar to those of other Be systems, with no clear-cut separation between them. One of the new systems is a candidate for a rare case of quadruple system involving a Be star. Five additional \gc\ stars display velocity variations compatible with the presence of companions, but no orbital solution could yet be formally established for them hence they only receive the status of ``binary candidate''.
\end{abstract}


\section{Introduction}
The Be star category was first described nearly 150 years ago with the discovery of emission lines in \gc, the central star of Cassiopeia's ``W'' \citep{sec66}. These lines are now thought to arise in a viscuous decretion disk (for a review, see \citealt{riv13}). However, the prototype star, \gc, was found to be quite atypical amongst its defining class. Indeed, it emits very bright and extremely hard X-rays with a luminosity intermediate between that of ``normal'' massive stars and that of high-mass X-ray binaries (for a review, see \citealt{smi16}). Over the last two decades, it was discovered that \gc\ is not a lone bird, hence the definition of a ``\gc '' category. Currently, 25 Be stars are known to display similar X-ray properties (see \citealt{naz18} and \citealt{naz203new}): such \gc\ objects thus represent a non-negligible fraction of the classical Be population.

The mechanism behind the peculiar X-ray emission of \gc\ analogs is still debated, with a handful scenarios having been proposed. The first one involves accretion onto a compact companion (white dwarf, WD, \citealt{mur86}; neutron star, NS, in a propeller regime, \citealt{pos17}). A second one imagines the collision between the stellar wind of a hot, non-degenerate, companion and the Be disk \citep{lan20}. A third one considers star-disk interactions involving localized magnetic fields, generated by disk instabilities \citep{smi98,smi99}, and stellar fields arising from a subsurface equatorial convective zone \citep{can09,mot15}. A crucial difference between these scenarios is the presence (and nature) of companions. If either the accretion scenario or the wind collision idea is correct, {\it all} \gc\ stars must have a companion. In the first case, this companion would be a {\it compact} object with a mass typical of either a white dwarf or a neutron star. In the second case, the companion would be a stripped Helium star whose mass would span the range expected for such objects (typically a few $M_{\odot}$, the exact value being linked to the Be star mass, see \citealt{sha21}). For the last scenario, companions are possible, but their presence is certainly not a requirement since X-ray production is not linked to the companion. Consequently, assessing the systematic presence of {\it close} companions to \gc\ stars constitutes a crucial step towards constraining the nature of their peculiar X-ray emission.

Furthermore, the binarity status is of particular interest in the context of evolutionary models \citep{sha14,sha21}. Indeed, in the first scenario, \gc\ stars could fill the theoretically-predicted population of Be+WD systems \citep{rag01}, which should largely outnumber Be+NS X-ray binaries although only a handful of cases are known. Alternatively, the detection of neutron star companions would lead to an increase in the number of NS+OB systems, also impacting on evolutionary theories. Finally, the presence of non-degenerate companions would lead to constraints on the origin of Be stars themselves, in addition to information on stellar evolution. Indeed, the Helium stars have lost their envelope after a mass-transfer episode and, in such an event, angular momentum is also transfered: the associated spin-up would then explain the fast rotation of Be stars \citep{sha14}. 

Amongst the 25 known \gc\ stars, two are known binaries: \gc\ itself \citep[e.g.][]{nem12,smi12} and $\pi$\,Aqr \citep{bjo02,naz19piaqr}. The velocity variations of a third object, HD\,45314, were examined by \citet{rau18} who found no evidence for periodic velocity variations with amplitude $>10$\,\kms. For the other \gc\ stars, systematic monitorings have not yet been done. Thanks to dedicated pilot programmes, we begin this endeavour using TIGRE and CARMENES for seven northern targets and UVES for nine southern targets (see further details below). The six remaining \gc\ stars are too faint for a detailed study with current facilities. Section 2 presents the obtained observations as well as the velocity mesurement methods. Section 3 examines the results obtained, with a specific focus on the new detections. Section 4 discusses the results, while the last section summarizes our findings.

\subsection{The binarity of Be stars}
Before examining our results and their impact on the debate regarding the \gc\ phenomenon, the more general question of multiplicity in the overall Be category should also be presented. It has been known for decades that some Be stars are binaries. Indeed, tens of them show up brightly at X-ray or gamma-ray wavelengths \citep[and references therein]{wal15}. In such cases, the Be star is paired with a neutron star (NS) which accretes material from its companion, explaining the intense high-energy emission. Of course, this bright emission makes them detectable from afar and the sheer number of such systems may induce a false idea of great abundance. Considering volume-limited samples, such X-ray binaries actually represent only a small fraction of the Be population. Theoretical simulations by \citet{rag01} indeed predict 70\% of white dwarf (WD) companions, 20\% of stripped, He-star companions, and only 10\% of NS companions in Be binaries. More recently, \citet{sha14} provided additional support to a low fraction of NS companions as they calculated that there should be 50 Be+NS systems for each Be+BH case, but a thousand other Be binary systems for each Be+NS case. This thousand should be equally split between Be stars with WD companions and Be stars with stripped, He-star companions. Furthermore, binary properties may differ between these different cases. For example, both \citet{rag01} and \citet{sha14} expect a large range of eccentricities for NS companions, depending on the supernova nature and kick. In contrast, more circular systems are generally expected when mass transfer has occurred but no supernova has taken place, as for WD and He-star companions (although some simulations suggest non-zero eccentricities, see \citealt{sep09,dos16}). Regarding orbital periods, \citet{rag01} favor larger values for NS (100--3000\,d) then for WD and He-star companions (20--300\,d) while \citet{sha14} found similar ranges (20--2000\,d) in all cases. \citet{sha14,sha21} also expected a correlation between the mass of the Be star and the mass of the He-star: the stripped companion should have a mass of 0.5--3\,M$_{\odot}$ for a Be mass of 6--15\,M$_{\odot}$. Furthermore, for that range of the Be mass, the period distribution appeared rather flat.

The difficulty of detecting these other pairs may strongly bias the observed incidence of such companions, however. For example, the expected large population of Be stars with WD companions has not been found yet. Only a few cases were reported, e.g. through the detection of Be stars associated to supersoft X-ray sources (e.g. \citealt{cra18,ken21}). In parallel, the UV radiation of hot, stripped He-stars allowed to detect such companions, but only for a few Be stars as the detection is here also very challenging \citep{wan18,wan21}. Finally, while orbits could be established for some Be binaries (see Sect. 4 below), the companion rarely (if at all!) seems to be a main-sequence star with similar mass. For example, \citet{bod20} mentioned a clear lack of massive companions (with mass ratios $>0.2$, i.e. A- and B-type) to early-type Be stars, compared to what is found in other B-type stars. 

Thus, the overall picture that emerges from previous research is that Be stars may be found in binary systems, often (if not always) with an evolved companion. The following question then concerns the actual fraction of Be binaries amongst the Be population. This is often debated in the context of evolutionary models and the emergence of Be signatures. Indeed, the Be stars are known to be fast rotators, and this is considered as an essential property to achieve the Be phenomenon (although probably not the sole one as some fast rotating B-stars are not of the Be type). There are two ways to obtain such a fast rotation. On the one hand, one may consider B-type stars as being born with a range of rotational velocities and only the fastest rotators turning into Be stars (with or without the need of a further spin-up during main sequence, see \citealt{zor97} vs. \citealt{eks08}). In this case, binarity brings ``no extra clue on the origin of the Be phenomenon'' \citep{baa92} as it constitutes an unrelated factor. On the other hand, one may consider the fast rotation as acquired after a transfer of angular momentum in an interacting binary system. A first proposal had been that Be stars correspond to systems {\it currenly} undergoing mass-transfer, with the Be disk being an accretion feature (e.g. \citealt{kri75}). This scenario was later discarded as such a phase is short-lived and the necessary mass-losing companions (cool giants) were not detected in most Be stars. Therefore, the binary idea was restricted to post mass-transfer systems. In such a case, the donor star would then appear as a stripped star which could end up as a WD or NS. Theoretical estimates of the relative importance of one channel over the other have been done, with widely different results. For example, \citet{van97} favored little contribution (maximum 20\% of all Be stars) of the binary channel, whereas \citet{pol91} or \citet{sha14} considered it much more dominant (40--60\% of Be stars for the former, $\sim$100\% for the latter).

To estimate the overall binarity fraction of Be stars, specific observational studies have also been undertaken. \citet{bou18} estimated the fraction of runaways amongst Be stars and found that it agrees with expectations for post mass-transfer systems, suggesting the binary channel to be dominant. \citet{kle19} examined the SED of a sample of Be stars. In these stars, the SED combines the usual stellar photospheric emission with the emission from the cooler disk. \citet{kle19} found SED turndowns at radio wavelengths for all stars with sufficient wavelength coverage. They interpreted that result as evidence for disk truncation by companions, suggesting that the vast majority of Be stars possess companions and are born through the mass-transfer channel. Finally, \citet{has21} could reconcile the fraction of Be stars observed in clusters with predictions of binary models. However, they noted that, ``if all Be stars are binary interaction products, somewhat extreme assumptions must be realised'' for the match to occur. In particular, all stars are binaries in their model and, whatever the stellar and orbital properties, each system undergoes a fully non-conservative mass-transfer immediately at the end of the main sequence, which necessarily leads to the formation of a Be star.

\section{Observations and data reduction}
The list of observed stars and their basic properties is provided in Appendix A (Table \ref{liststar}).
  
\subsection{CARMENES}
Four northern stars (V782\,Cas=HD\,12882, V810\,Cas=HD\,220058, V2156\,Cyg, and SAO\,49725) were observed with the CARMENES \citep[Calar Alto high-Resolution search for M dwarfs with Exoearths with near-IR and optical Echelle Spectrographs,][]{CARMENES} instrument. All data were obtained in service mode via the OPTICON common time allocation process (programs ID 19B/001+21B/017) at a rate of about one observation every two weeks in August--November 2019 and July-October 2021. Installed on the 3.5\,m telescope of the Centro Astron\'omico Hispano Alem\'an at Calar Alto Observatory (Spain), CARMENES records the stellar spectra thanks to two separate echelle spectrographs, one for red wavelengths (5200--9600\,\AA, $R = 94\,600$) and one for the near-IR (0.96--1.71\,$\mu$m, $R = 80\,400$). The spectra were reduced using the CARMENES pipeline \citep{cab16}. The individual spectra were taken with exposure times between 1 and 15\,min depending on star brightness and weather conditions. If obtained the same night, they were combined to improve signal-to-noise ratios. The final spectra (about ten for each star) display typical signal-to-noise ratios of 150--200. Within IRAF, a further correction was made for eliminating telluric lines around H$\alpha$ and He\,{\sc i}\,$\lambda$\,5876\AA\ in the optical (template of \citealt{hin00}), as well as around 1$\mu$m and 1.28$\mu$m in the near-IR (template as in \citealt{naz19cyg12}). Sky emission lines near 1.56$\mu$m were also eliminated, using the data from the sky fiber. As a last step, the spectra were normalized over limited wavelength windows using splines of low order. 

\subsection{TIGRE}
Four bright northern/equatorial stars (V782\,Cas=HD\,12882, V558\,Lyr=HD\,183362, HD\,44458=FR\,CMa, and HD\,45995) were observed between November 2018 and {\bf October} 2021 with the fully robotic 1.2\,m TIGRE telescope located in central Mexico \citep{Schmitt}. A typical bimonthly cadence was used over the visibility period, weather permitting: in total, we collected 33 spectra of V782\,Cas, 16 spectra of V558\,Lyr, 20 spectra of HD\,44458, and 19 spectra of HD\,45995. As a complement to the CARMENES observations, three spectra of V810\,Cas and V2156\,Cyg, plus one of SAO\,49725 were also collected.

TIGRE is equipped with the HEROS echelle spectrograph (3800--5700\,\AA\ + 5800--8800\,\AA, $R\sim20\,000$). The data reduction was performed with the dedicated TIGRE/HEROS reduction pipeline \citep{Mittag}. Absorption by telluric lines was corrected within IRAF using the template of \citet{hin00} around He\,{\sc i}\,$\lambda$\,5876\AA\ and H$\alpha$. As a last step, the high-resolution spectra were normalized over limited wavelength windows using splines of low order. Typical exposure times were between 5 and 30\,min, depending on star brightness and weather conditions; typical signal-to-noise ratios (after box smoothing over 7 pixels) were about 200. Five spectra of V782\,Cas showing very low signal-to-noise ratios were discarded, as well as another one of the same star showing a deviating velocity for the interstellar Ca\,{\sc ii}\,3934\,\AA\ line.

\subsection{UVES}
Nine southern stars (HD\,90563, HD\,110432=BZ\,Cru, HD\,119682, V767\,Cen=HD\,120991, CQ\,Cir=HD\,130437, HD\,157832=V750\,Ara, HD\,161103=V3892\,Sgr, V771\,Sgr=HD\,162718, and HD\,316568) were observed with the Ultraviolet and Visual Echelle Spectrograph (UVES, \citealt{dek00}), installed on the second Unit Telescope (UT2) at the Cerro Paranal ESO Observatory, for our ESO program ID 105.204D. Spectra were taken on five dates between September 2020 and March 2021, with at least 10 days between successive exposures. UVES was used in dichroic mode, allowing simultaneous access to the 3300--4560\,\AA\ and 4730--6830\,\AA\ regions (with a 80\AA\ gap near $\sim$5800\AA). Exposure times ranged between 0.25 and 30\,min, depending on star brightness and weather conditions, leading to typical signal-to-noise ratios of 100--150 in the blue and 200--250 in the red. The slit width was 0.4\arcsec\ in the blue and 0.3\arcsec\ in the red, leading to $R\sim 70\,000$ and 100\,000, respectively. The data were reduced in a standard way by ESO pipeline. Telluric lines were corrected near He\,{\sc i}\,$\lambda$\,5876\AA\ and H$\alpha$ in the same way as for CARMENES and TIGRE spectra. Finally, the spectra were normalized over the same set of continuum windows using polynomials of low order. Spectra were taken in service mode, but were not always checked for saturation in H$\alpha$: only some saturated observations were repeated with a reduced exposure time - fortunately, the other lines are unaffected by this problem.
  
\subsection{\te}
For V782\,Cas, additional photometric data have been obtained by \te\ since those presented in \citet{naz20tess} hence we re-extracted the whole dataset. The 30\,min lightcurves of both sectors (18 and 25) were derived through aperture photometry using the Python package Lightkurve\footnote{https://docs.lightkurve.org/}. On image cutouts of 50$\times$50 pixels, we defined a source mask from pixels presenting fluxes well above the median flux (10 or 25 Median Absolute Deviation over it) and a background mask from pixels with fluxes below the median flux. The background contamination was then corrected using a principal component analysis with 5 components. All data points with errors larger than the mean of the errors plus three times their 1$\sigma$ dispersion were discarded. The \te\ fluxes were converted into magnitudes using $mag=-2.5\times \log(flux)$, the mean magnitude in each sector was subtracted, and then the data were combined into a final \te\ lightcurve\footnote{Note that this new reduction of the \te\ data of V782\,Cas in Sector 18 produce a lightcurve fully agreeing with that reported in \citet{naz20tess}.}.

\begin{figure*}
  \begin{center}
    \includegraphics[width=18cm,bb=30 510 575 700, clip]{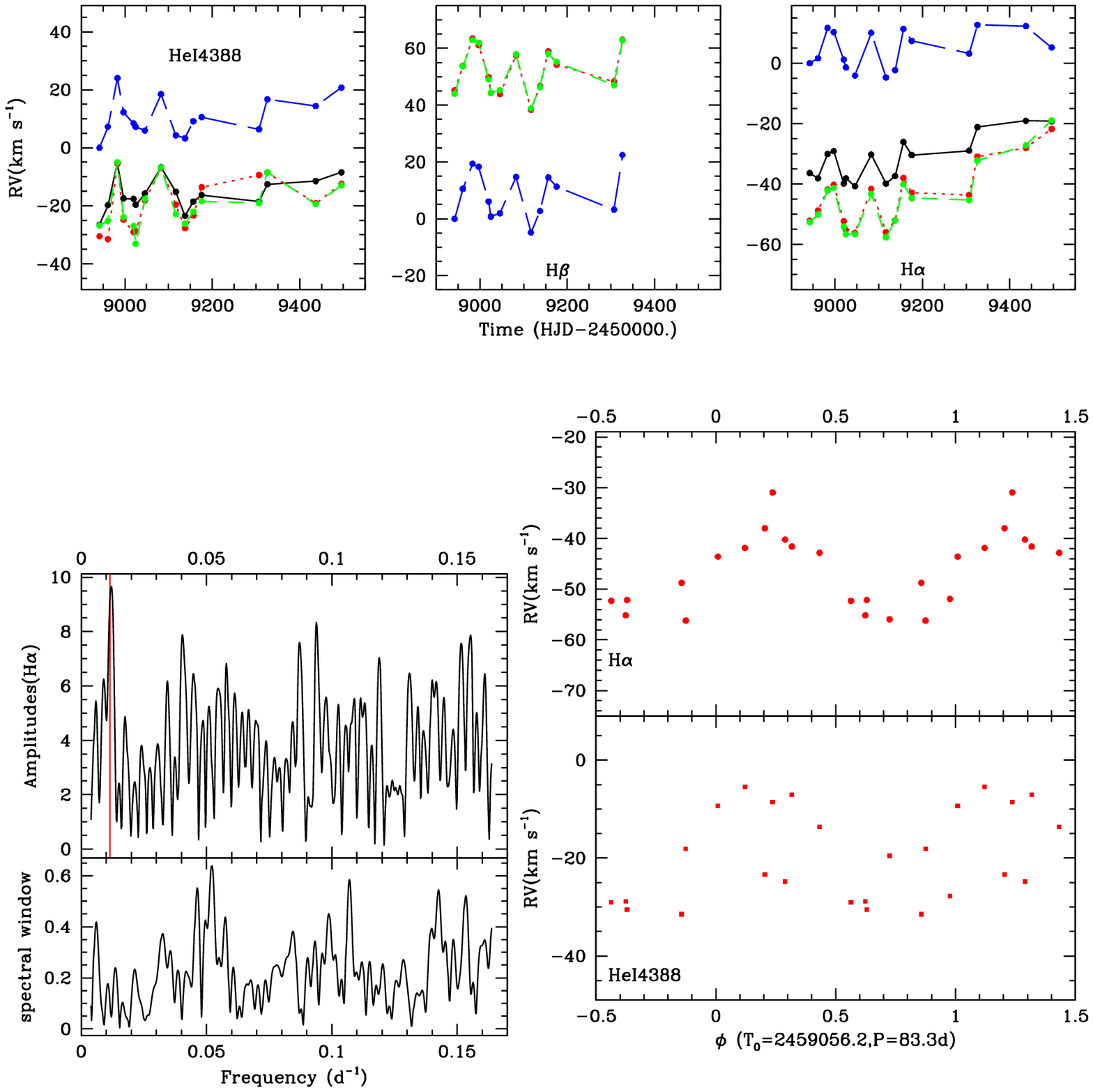}
  \end{center}
  \caption{Radial velocities measured for V558\,Lyr on He\,{\sc i}\,4388\,\AA\ (left), H$\beta$ (middle) and H$\alpha$ (right) lines using the moment method (black points and solid line), the mirror method (red points and dotted line), the double Gaussian method (green points and dashed line), and the correlation method (blue points and long dashed line). Note that the mix of absorption and emission in H$\beta$ makes the first order moments unreliable hence they are not shown. }
\label{v558rv}
\end{figure*}

\subsection{Measuring velocities of Be stars}
Our aim is to detect binary motion through radial velocity measurements. However, several phenomena can lead to velocity changes, fortunately with different properties. Binary motion should affect all lines of a given star in the same way, while pulsations or disk evolution will show up differently depending on line. This allows us to disentangle these effects: only stars showing velocity shifts which are coherent from line to line will be considered as ``binary candidates''. If a period and a full orbital solution can be calculated, then the status becomes secure and the ``candidate'' is dropped.

However, measuring radial velocities ($RVs$) in Be stars is not a simple endeavour, as their lines are broad, with few and faint photospheric lines being present while emission lines may be strong but display variable profiles due to disk build-up or dissipation events. To characterize the lines, we therefore (1) carefully selected them and (2) used several methods to constrain their $RVs$.

To get reliable $RVs$, strong lines must be studied. For CARMENES spectra, we chose lines from hydrogen (H$\alpha$, Paschen lines at 0.8598, 0.8750, 1.00, and 1.28$\mu$m, plus the Brackett line at 1.56$\mu$m) as well as from He\,{\sc i}\,5876\,\AA\ and Fe\,{\sc ii}\,9998\,\AA. For TIGRE and UVES data, we focused on lines from hydrogen (H8, H$\beta$, H$\alpha$) and helium (He\,{\sc i}\,4026,4143,4388,4471,5876,6678\,\AA). Of those sets, H8 is the only hydrogen line with a dominant absorbing component. The line is however very broad and quite noisy, making it unreliable. Lines of He\,{\sc i}\,5876,6678\,\AA\ usually show a mix of absorption and emission, while bluer He\,{\sc i} lines appear dominated by absorption. In addition to stellar lines, we also studied interstellar lines (Ca\,{\sc ii}\,3934\,\AA, Na\,{\sc i}\,5895\,\AA, DIB at 5780\AA) to get an idea of the precision of the $RV$ measurement methods.

Four methods were used to measure the $RVs$. The first one calculates the first-order moment $M_1$ using $M_1=\sum (F_i-1)\times v_i /\sum (F_i-1)$ \citep{naz19piaqr}, where $v_i$ the velocity of the ith spectral bin and $F_i$ the normalized flux at the ith spectral bin. This method appears quite robust in presence of noise since it considers the whole line profile. For the same reason, however, it is also sensitive to profile changes. This may become particularly problematic for lines strongly affected by disk emission (e.g. H$\alpha$ emissions) as the line core, linked to the outer part of the disk, is often quite variable. In addition, it provides erratic results if the equivalent width is small (because the denominator is then close to zero), i.e. it does not work well for lines mixing absorption and emission of similar amplitudes.

The second method is the correlation method. It compares one spectrum (taken as reference) to all others for a set of velocity shifts. The best shift is found by ways of a $\chi^2$ minimization and a parabolic fit to the minimum $\chi^2$ value and its neighbours is then made to find the final value of the relative velocity. Like the previous method, it uses the whole line profile hence may become biased if the profile strongly changes. However, it does not have any problem with mixed absorption/emission profiles.

The next two methods put emphasis on the wings of the line profile, avoiding the line core. They both require the lines to have quite large amplitudes, compared to the noise. The mirror method compares, for several velocity shifts, the blue wing to the mirrored red wing (i.e. after reversing velocities, \citealt{nem12} and references therein). Only the wings between approximately 20\% and 60\% of the maximum amplitude were considered in this calculation, as done for $\gamma$\,Cas \citep{nem12}. The least difference between the wings is searched for and a parabolic fit to the minimum $\chi^2$ value and its neighbours is made to get the final velocity value. While it avoids the core, this method requires the wings to be spread over a number of pixels: it is thus difficult to use on steep lines such as interstellar lines.

The double Gaussian technique correlates the line profile to the function $G(v)=exp[-(v-a)^2/2\sigma^2]-exp[-(v+a)^2/2\sigma^2]$ where $v$ is the velocity, $a$ the center of the positive Gaussian function ($-a$ is used for the negative one), and $\sigma$ the standard deviation of both Gaussian functions \citep[based on \citealt{sha86}]{smi12}. The $RV$ corresponds to the shift at which the correlation reaches zero. In each case, we adjust parameter $a$ to put maximum weight where the line reaches half maximum (i.e. it is a bisector value at half amplitude). The width $\sigma$ was chosen to be 10\,\kms\ (for CARMENES and UVES) or 15\,\kms\ (for TIGRE) for stellar lines and 5\,\kms\ for the narrower interstellar lines. 

Because these four methods are sensitive to different aspects of the line profile and because the line shape (e.g. asymmetric profiles) may skew the results of one method but not another, the absolute velocity values found by these four methods may not agree. However, for simple cases (e.g. no strong changes of line core), their relative variations should agree. Note that we determined width and skewness of each line in each star, in order to check for the presence of large line profile changes. Except for the last spectra of V558\,Lyr, the northern targets displayed relatively stable profiles, while this was much less the case for southern stars. Nevertheless, aside from interstellar lines, the H$\alpha$ line appears as the most stable line in the vast majority of cases.

\section{Results}

Whatever the method, the velocities derived for the narrow interstellar lines (Ca\,{\sc ii} and/or Na\,{\sc i}) yield dispersions of $\sim$1\,\kms\ for TIGRE spectra, $\sim$0.1\,\kms\ for CARMENES spectra, and $\sim$0.15\,\kms\ for UVES spectra. This corresponds well to the expected resolution and wavelength calibration limits. For the broader and very shallow DIB recorded on CARMENES spectra, the dispersion however rises to $\sim$1\,\kms. Although the stellar lines are less shallow, the actual uncertainty on their velocities may thus be somewhat larger than for narrow lines.

Four stars were repeatedly observed with TIGRE, thereby allowing a finer study. Indeed, whenever recurring variations are found, an orbital solution could be derived since there were enough spectra. The specific results on these four stars are thus discussed in separate subsections (Sect. 3.1--4). The CARMENES+TIGRE observations of three additional targets are fewer in number but provided first hints towards orbital solutions, presented in Sect. 3.5. Finally, the UVES datasets of other \gc\ stars only allow for a preliminary assessment, hence they are discussed in the last subsection (Sect. 3.6). For all stars, the Appendix provides the full list of measured velocities (Tables \ref{rv782b} to \ref{rv810}). Note that all orbital solutions derived below use simple sinusoids, i.e. consider zero eccentricity: this comes from (1) the small eccentricity revealed by the phase-folded velocities and (2) the quality of the data (limited number of observations, velocity uncertainties) especially in view of the small orbital motions detected. 

\subsection{V558\,Lyr}

Up to now, no companion was reported in the literature \citep[ although only 3 spectra taken the same day were studied in the latter case]{abt84,bec15}. However, our radial velocities of V558\,Lyr clearly display recurrent variations, whatever the method used to measure them (Fig. \ref{v558rv}). These variations are best seen for the H$\beta$ and H$\alpha$ lines. The very broad H8 line shows variations of an overall similar character but much noisier, rendering its results unusable. For the He\,{\sc i} lines, which are in absorption, the same wiggle is seen but superimposed on an increasing trend for all lines but He\,{\sc i}\,4388\,\AA. The fact that the motion truly seems to affect the whole spectrum indicates without ambiguity the signature of binary motion. No significant trace of lines from the companion are seen in our spectra (e.g. Calcium triplet in the near-IR), unfortunately, so that V558\,Lyr can only be classified as SB1 for the moment.

In the most recent spectra, line profile changes are detected. For example, the H$\beta$ line showed a double-peaked line profile with a strongly dominating red peak until April 2021. Afterwards, in the last two spectra, the peaks become more equal, with the blue peak finally becoming the strongest one. Similar blue/red variations are detected in other lines. To avoid any confusion in the nature of the velocity shifts, we discarded the last two spectra from further consideration in the orbital derivation. 

\begin{figure}
  \begin{center}
    \includegraphics[width=8cm,bb=30 150 285 425, clip]{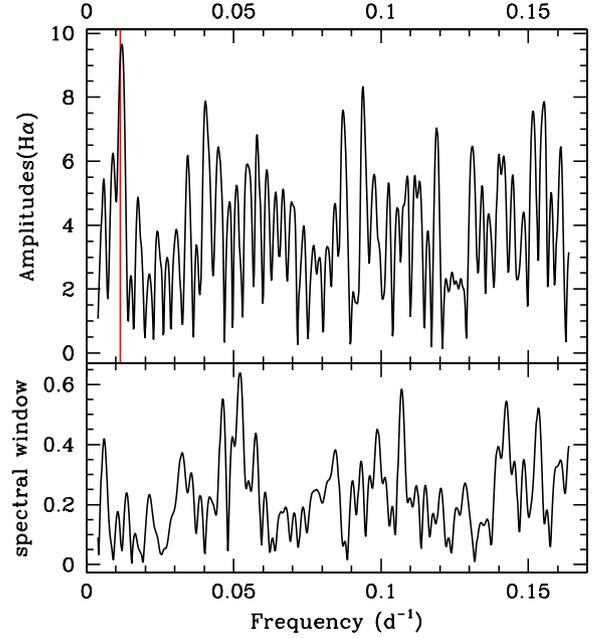}
  \end{center}
  \caption{Periodogram derived for H$\alpha$ velocities of V558\,Lyr measured with the mirror method by the modified Fourier algorithm, along with its spectral window. A vertical line indicates the proposed orbital period.}
\label{v558fou}
\end{figure}

\begin{figure}
  \begin{center}
    \includegraphics[width=8cm,bb=290 150 575 500, clip]{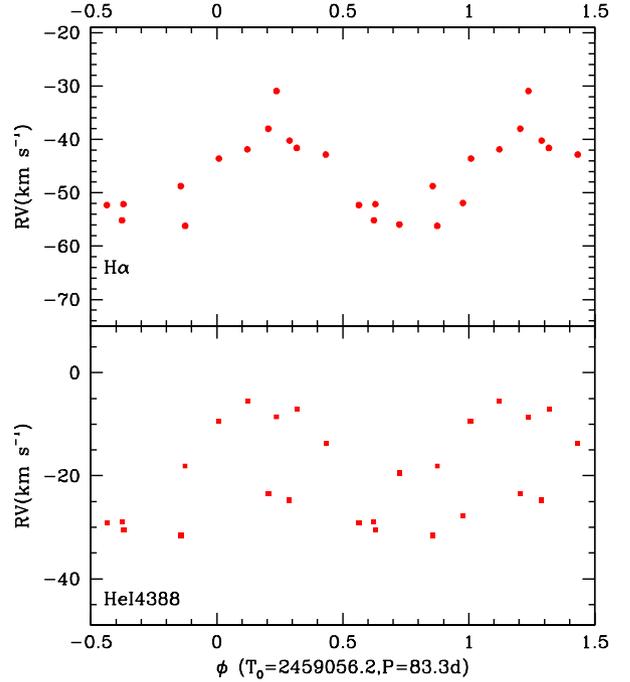}
  \end{center}
  \caption{Velocities measured for V558\,Lyr with the mirror method for He\,{\sc i}\,4388\,\AA\ (bottom) and H$\alpha$ (top) folded with the best-fit ephemeris. }
\label{v558phi}
\end{figure}

We performed a period search on H$\alpha$ velocities (derived with the mirror method), using several methods (modified Fourier periodogram \citealt{hmm,gos01,zec09}, AOV \citealt{sch89}, conditional entropy \citealt{cin99,cin99b,gra13}). A peak is clearly detected at 83.3$\pm$1.8\,d in all methods (e.g. Fig. \ref{v558fou}). It appears at 2.6 times the mean level of the modified Fourier periodogram, but the fact that the modulation is seen on several lines leaves little doubt on the binary nature. Only with more data (there are only 14 points after all!) can the peak better stand out and a more precise period value be found. A $\chi^2$ fit yields a best-fit sinusoid with parameters: $T_0=2\,459\,056.2\pm1.4$ (conjunction with the Be primary in front, Fig. \ref{v558phi}) and $K=8.2\pm1.1$\,\kms. The H$\alpha$ velocity dispersion passes from an initial 7\,\kms\ to only 3\,\kms\ once that best-fit sinusoid is taken out. 

The best-fit parameters yield a mass function of 0.0048$\pm$0.0019\,$M_{\odot}$. Considering the Be star (spectral type B3V) to have a mass of 8\,$M_{\odot}$ \citep{vie17} and the system's inclination to be 60--90$^{\circ}$, the companion should then have a mass of 0.7--0.8$\pm$0.1\,$M_{\odot}$. The values of period and velocity amplitude derived for the Be star in V558\,Lyr are similar to those found for HR\,2142 \citep{pet16} although the type of Be primary of the latter system (B1.5IV-V) appears earlier than for the Be star in V558\,Lyr. It may be noted in this context that no hint of a ``hot'' companion (i.e. stripped Helium star) was found for V558\,Lyr by \citet{wan18}. However, this was also the case for HR\,2142: its companion is too faint hence its presence is not easily detected. Therefore, the possibility of a hot companion to V558\,Lyr cannot be dismissed on that basis. 

\begin{figure}
  \begin{center}
    \includegraphics[width=8.5cm,bb=20 170 550 720,clip]{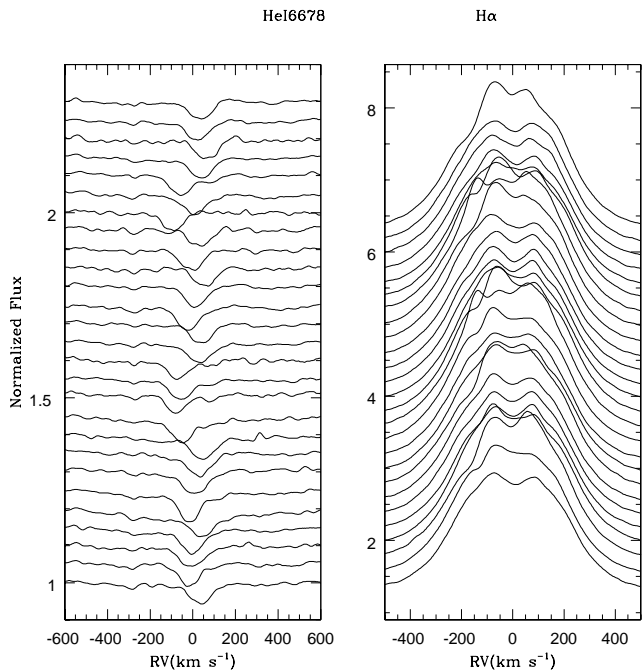}
  \end{center}
  \caption{Evolution with time of the He\,{\sc i}\,6678\,\AA\ (left) and H$\alpha$ (right) line profiles observed for V782\,Cas with TIGRE. Time is running downwards. }
\label{v782prof}
\end{figure}

\begin{figure*}
  \begin{center}
    \includegraphics[width=18cm,bb=25 515 590 700, clip]{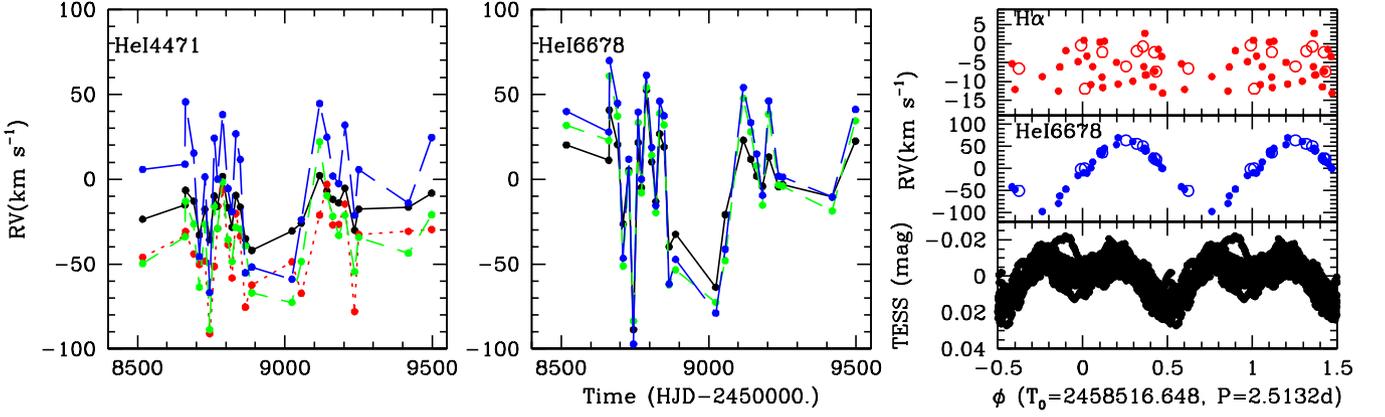}
  \end{center}
  \caption{{\it Left and Middle:} TIGRE radial velocities measured for V782\,Cas on He\,{\sc i}\,4471,6678\,\AA\ (left and middle) using the moment method (black points and solid line), the mirror method (red points and dotted line), the double Gaussian method (green points and dashed line), and the correlation method (blue points and long dashed line). Note that the mirror method sometimes fails for the He\,{\sc i}\,6678\,\AA\ because of the narrowness of the line, hence its associated velocities are not shown. {\it Right:} \te\ photometry, He\,{\sc i} velocities (derived by the correlation method) and H$\alpha$ velocities (measured with the mirror method) folded with the best-fit 2.5\,d ephemeris of V782\,Cas. For $RVs$, filled points correspond to TIGRE data and the open circles to CARMENES data - note their good agreement.}
\label{v782rv}
\end{figure*}

\begin{figure}
  \begin{center}
    \includegraphics[width=8.5cm,bb=55 450 570 690, clip]{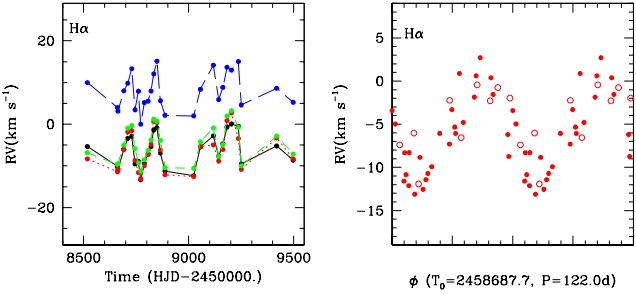}
  \end{center}
  \caption{{\it Left:} Same as left and middle panels of Fig. \ref{v782rv} but for the H$\alpha$ velocities. {\it Right:} H$\alpha$ velocities (measured with the mirror method) folded with the best-fit 122\,d ephemeris. Filled points correspond to TIGRE data and the open circles to CARMENES data.}
\label{v782phi}
\end{figure}

\begin{figure}
  \begin{center}
    \includegraphics[width=8.5cm,bb=50 150 550 750,clip]{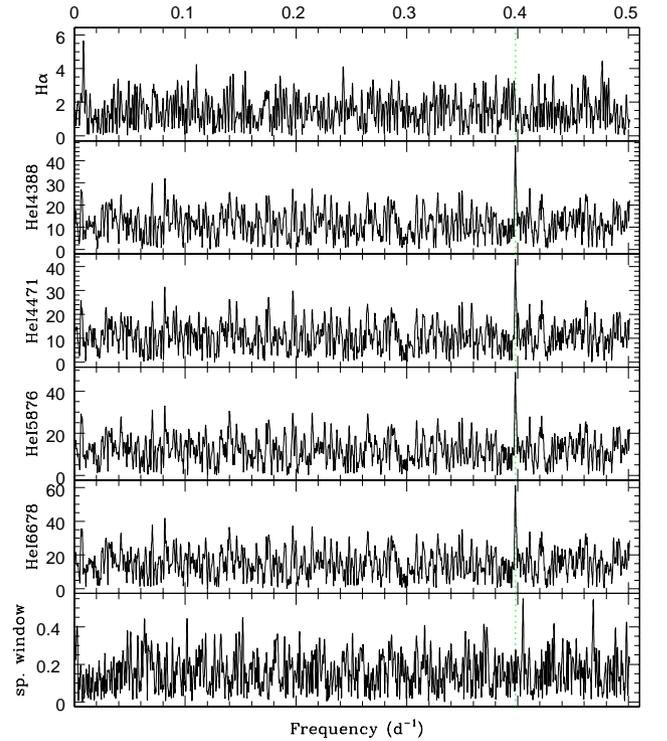}
  \end{center}
  \caption{Periodogram derived for He\,{\sc i} velocities (obtained by the correlation method on TIGRE data) and for H$\alpha$ velocities (measured with the mirror method on TIGRE data) of V782\,Cas by the modified Fourier algorithm, along with its spectral window. The dotted green line indicates the photometric frequency \citep{naz20tess}.}
\label{v782fou}
\end{figure}

\subsection{V782\,Cas}
\subsubsection{Discovery of a double periodicity}

On both CARMENES and TIGRE spectra, a very distinct and rather narrow absorption feature is clearly seen in the He\,{\sc i}\,5876,6678\,\AA\ lines of V782\,Cas. That component covers $\sim$220\,\kms\ and has $FWHM\sim 130$\,\kms. It presents substantial shifts from one observation to the next (left panel of Fig. \ref{v782prof}). The He\,{\sc i} lines at lower wavelengths appear broader (covering $\sim700$\,\kms), with the narrow component buried in the overall profile. The H$\alpha$ profile also presents variations, but of a different phenomenology (right panel of Fig. \ref{v782prof}).

The difference in behaviour is contrasted in Figs. \ref{v782rv} and \ref{v782phi}: the He\,{\sc i} velocities vary by a large amount (up to 150\,\kms, Fig. \ref{v782rv}) on short timescales whereas the H$\alpha$ velocities rather display a smooth, shallower (15\,\kms\ amplitude), and longer-term modulation (Fig. \ref{v782phi}). Performing period searches on the set of TIGRE velocities (because they are more numerous) clearly confirm the difference (Fig. \ref{v782fou}), with periods of 2.5132$\pm$0.0006\,d found for He\,{\sc i} lines and 122.0$\pm$1.5\,d found for H$\alpha$. The peaks reach amplitudes of $\sim$4 times the mean level of their respective periodograms, and it is important to note that no trace of the 2.5\,d signal is seen in H$\alpha$ periodogram and of the 122\,d signal in He\,{\sc i} periodogram (Fig. \ref{v782fou}). 

The shorter period is reminiscent of the photometric periodicity reported by \citet{lab17} and \citet{naz20tess}. In the latter reference, those photometric changes were interpreted as signatures from eclipses in a system unrelated to the Be star. This is reinforced by the fact that the $RUWE$ (Renormalised Unit Weight Error) parameter in Gaia-DR3 catalog is 1.87, a large value being generally indicative of a very close optical companion\footnote{https://gea.esac.esa.int/archive/documentation/GDR2/ Gaia\_archive/chap\_datamodel/sec\_dm\_main\_tables/ssec\_dm\_ruwe.html}. The recorded spectra would then be a superposition of the signatures of the Be system (a component which we call A) and of the lines from a binary (a component which we call B), see Table \ref{compv782t}. This helps explaining the weird line behaviour mentioned above. Most probably, the Be contribution to He\,{\sc i}\,5876,6678\,\AA\ is filled-in by emission, as often encountered in Be stars. With such an absorption/emission balance, the Be contribution becomes indistinguishable from the continuum, leaving only the contribution from the primary of the eclipsing system detectable. On the other hand, no (or little) emission contaminates the other He\,{\sc i} lines, allowing both the broad Be absorption and the narrower component of the eclipsing system to be detected.

The blue He\,{\sc i} lines therefore contain the contributions from two stars and the velocity measurement methods have different sensitivities to this mixing. The correlation method provides results clearly dominated by the narrow component, whereas the moment method rather yields a sort of average. Indeed, the velocity amplitude derived from moments appears to decrease when the Be absorption becomes more dominant (i.e. for blue He\,{\sc i} lines). Finally, wing methods have difficulties to provide sensible results, especially because of the rather large noise of blue He\,{\sc i} lines.

\begin{table}
  \caption{Estimated properties of the components in V782\,Cas. Note that the value of 57\% corresponds to the combined contribution of component A (i.e. Aa+Ab) to the total light. \label{compv782t}}
\begin{center}
  \begin{tabular}{lccc}
    \hline
    Name & sp. type & $M$ & light \\
         &          & (M$_{\odot}$)& fraction \\
    \hline
Aa & B2.5IIIe & 9    & 57\% \\
Ab &          & 0.65 &      \\
Ba & B2       & 9    & 35\% \\
Bb & B5       & 6    &  8\% \\
    \hline
  \end{tabular}
  \end{center}
\end{table}

\subsubsection{V782\,Cas\,B, the short-period binary}
We used the velocities of the He\,{\sc i} lines determined by the correlation method to determine a preliminary orbital solution. Averaging on all He\,{\sc i} lines, the best-fit sinusoid has an amplitude of 64$\pm$10\,\kms\ with $T_0=2\,458\,519.151\pm0.013$ (conjunction with the primary star in front). The velocity dispersion passes from an initial 40\,\kms\ to only 7\,\kms\ once that best-fit sinusoid is taken out. It may further be noted that forcing the fitting to H$\alpha$ velocities by a sinusoid with a period fixed to 2.5\,d yields no change in velocity dispersion, in line with the absence of a peak at that frequency in the periodogram (see also right panel of Fig. \ref{v782rv}). Since He\,{\sc i} lines are detected for the primary of the short-period binary despite the contribution of the bright Be star and since the He\,{\sc i} lines reach their maximum strength at spectral type B2, we conclude that the brightest component of this binary is likely near spectral type B2.  

\begin{table}
  \caption{SB1 orbital solution of the short-period binary in V782\,Cas as derived from the average $RVs$ of the He\,{\sc i} lines obtained via spectral disentangling. \label{solorbPshort}}
\begin{center}
  \begin{tabular}{c c}
    \hline
    Parameter & Value \\
    \hline
    P$_{\rm orb}$ (days) & 2.5132 \\
    $e$                & 0 \\
    $v_0$ (km\,s$^{-1}$) & $-15.5 \pm 1.2$ \\
    $K$ (km\,s$^{-1}$)   & $68.2 \pm 1.9$ \\
    $a\,\sin{i}$ (R$_{\odot}$) & $3.38 \pm 0.10$ \\
    $T_0$ (HJD$-$ 2\,450\,000) & $8516.648 \pm 0.009$ \\
    $f(m)$ (M$_{\odot}$) & $0.082 \pm 0.007$ \\
    \hline
  \end{tabular}
  \end{center}
\end{table}

\begin{figure}
  \begin{center}
      \includegraphics[width=8cm]{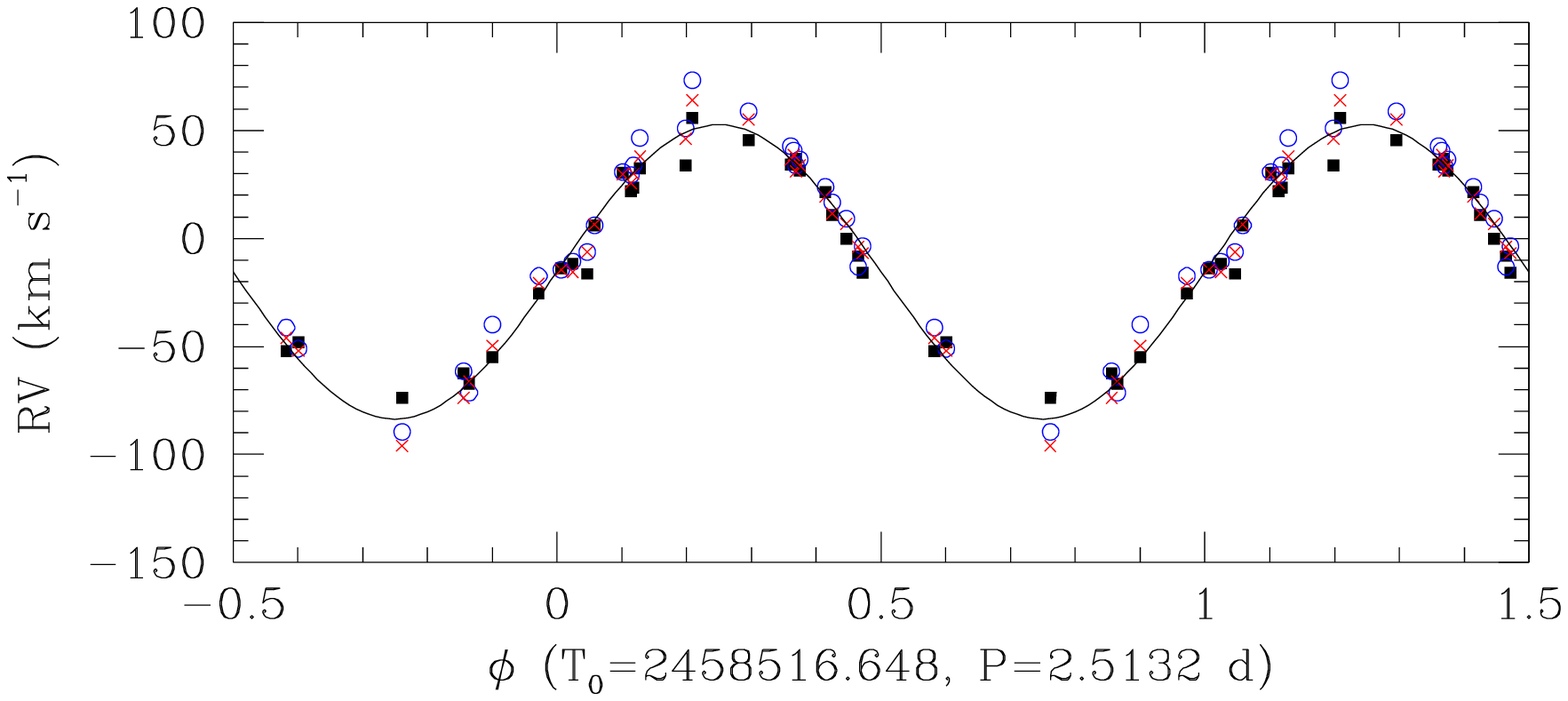}
      \includegraphics[width=8cm]{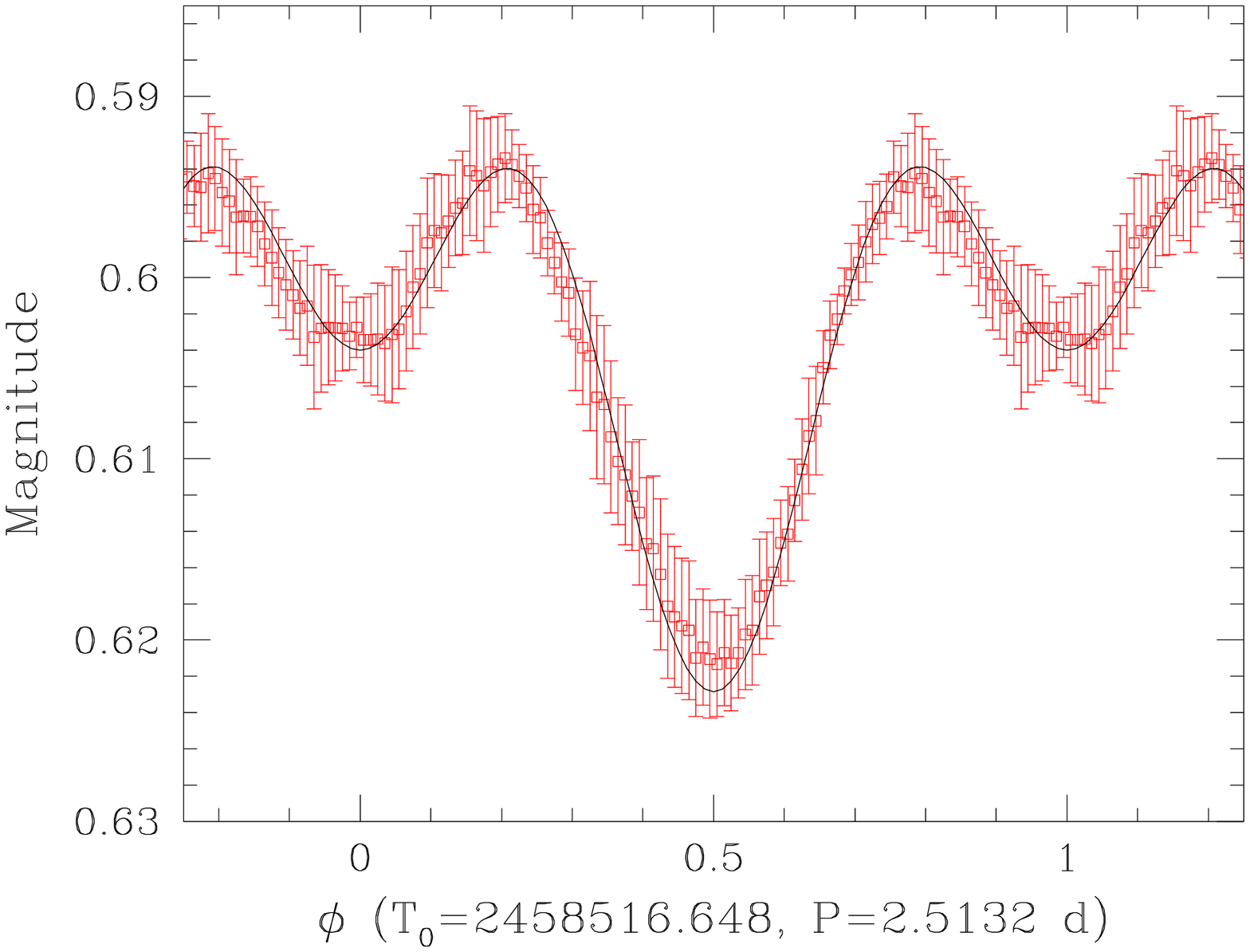}
  \end{center}
  \caption{{\it Top:} SB1 orbital solution obtained from the $RVs$ of the He\,{\sc i} absorption lines of V782\,Cas inferred via the spectral disentangling method. Filled squares, open circles and red crosses stand for the $RVs$ of the He\,{\sc i} 4713, 5876, and 6678\AA\ lines, respectively. {\it Bottom:} Best-fit {\tt Nightfall} binary model fit of the normal lightcurve of V782\,Cas built from the {\it TESS} data.\label{SB1HeI}}
\end{figure}

To further study this short-period binary, we have applied our spectral disentangling code \citep[e.g.][]{Rosu} based on the shift-and-add method of \citet{Gon06}. In most spectral regions, the method fails to distinguish the faint signature of the components of the short-period binary from the intrinsically variable contribution of the Be star. This is especially true for the regions around the Balmer lines. Significantly better results are obtained in the vicinity of the above-mentioned He\,{\sc i} lines, where we managed to reconstruct bits of the spectrum of the primary of the short-period binary. In the spectral disentangling procedure, the $RVs$ of the He\,{\sc i} lines inferred from the correlation method were used as input and updated $RVs$ were established by cross-correlation with a synthetic TLUSTY spectrum for a B2\,V star \citep{Lan03}. Using these updated $RVs$, we derived the SB1 orbital solution shown on top of Fig.\,\ref{SB1HeI} and in Table\,\ref{solorbPshort}.

Comparing the reconstructed He\,{\sc i} lines with TLUSTY spectra for different projected rotational velocities, we found the best agreement for $v_p\,\sin{i} = 75 \pm 10$\,km\,s$^{-1}$. Beside the three He\,{\sc i} lines used for the RV determination, the spectral disentangling unveiled absorption lines due to He\,{\sc i}\,$\lambda\lambda$\,4009, 4026, 4121, 4144, 4169, 4388, 4471\,\AA, C\,{\sc ii}\,$\lambda$\,4267\,\AA, Si\,{\sc iii}\,$\lambda\lambda$\,4552,4568\,\AA, and O\,{\sc ii}\,$\lambda$\,4649\,\AA. All these features were recovered with a strength consistent with a B2 spectral type classification diluted by a factor 0.28, i.e. the B2 component contributes 28\% of the total light of V782\,Cas. The CARMENES spectra are of higher spectral resolution, but are less numerous, rendering any attempt to apply the disentangling method to the He\,{\sc i} lines hopeless. Nevertheless, these data unveil a weak He\,{\sc i} $\lambda$\,5876 signature of the secondary of the short-period binary on three occasions. We deblended the primary and secondary contributions via a simultaneous fit of two Gaussians. Though the results of this fit must be considered with caution, the $RVs$ found for the primary nicely confirm the SB1 orbital solution inferred from the TIGRE data while the $RVs$ of the secondary suggest a value of $q= \frac{m_p}{m_s}$ between 1.3 and 1.7. Finally, the equivalent width of the secondary line appears to be about 1/4 of that of the primary. This suggests that the secondary He\,{\sc i} lines are diluted by about a factor 0.10--0.15, i.e. the component contributes 10--15\% of the total light of V782\,Cas. This rules out the possibility considered by \citet{naz20tess} that the short-period binary might correspond to one of the neighbouring sources resolved by {\it GAIA}, as these objects would contribute much less than $\sim$40\% to the combined light.

Since the {\it TESS} lightcurve of V782\,Cas indicates an eclipsing binary in an overcontact configuration \citep[see][and below]{naz20tess}, the stars in the short-period binary must be in co-rotation. We can then use the formula of \citet{Egg83} to express the radius of a star (over)filling its Roche lobe as a function of the mass-ratio $q$, and from there its projected rotational velocity:
\begin{equation}
  v_p\,\sin{i} = (G\,m_p)^{1/3}\,\left(\frac{2\,\pi}{P_{\rm orb}}\right)^{1/3}\,\frac{0.49\,q^{2/3}\,\left(1 + \frac{1}{q}\right)^{1/3}\,fill_p}{0.6\,q^{2/3} + \ln{(1 + q^{1/3})}}\,\sin{i}
\end{equation}
where $fill_p$ is the Roche lobe filling factor of the primary. With $P_{\rm orb} = 2.5132$\,days, this leads to
\begin{equation}
 v_p\,\sin{i} = 76.6\,fill_p\,\frac{q^{2/3}\,\left(1 + \frac{1}{q}\right)^{1/3}}{0.6\,q^{2/3} + \ln{(1 + q^{1/3})}}\,\left(\frac{m_p}{{\rm M}_{\odot}}\right)^{1/3}\,\sin{i}
\end{equation} 
where the numerical factor is expressed in km\,s$^{-1}$.

Comparing this expression with the observed $v_p\,\sin{i}$ then allows us to express a relation between orbital inclination and mass-ratio:
\begin{equation}
  \sin{i} = 0.978\,\frac{0.6\,q^{2/3} + \ln{(1 + q^{1/3})}}{fill_p\,q^{2/3}\,\left(1 + \frac{1}{q}\right)^{1/3}}\,\left(\frac{m_p}{{\rm M}_{\odot}}\right)^{-1/3}
  \label{rel1}
\end{equation} 

\begin{figure}
  \begin{center}
      \includegraphics[width=8cm]{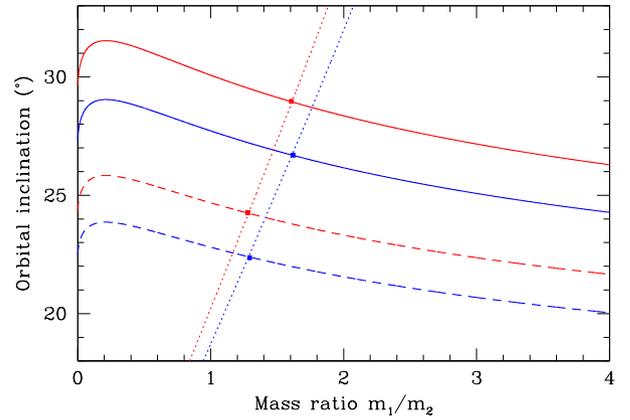}
  \end{center}
  \caption{Constraints on $i$ and $q$ from the projected rotational velocity (Eq.\,\ref{rel1}) and the mass function (Eq.\,\ref{rel2}) of the primary star of the short-period binary in V782\,Cas. The red and blue curves correspond to the results obtained for a primary mass of 8 or 10\,M$_{\odot}$. The continuous and dashed lines show the results for Eq.\,\ref{rel1} for $fill_p = 1.0$ and $fill_p = 1.2$, respectively. The dotted lines stand for the relation \ref{rel2} obtained from the mass function. \label{massratio}}
\end{figure}

This relation is displayed in Fig.\,\ref{massratio} for two different assumptions on $m_p$ (8 and 10\,M$_{\odot}$) and two different values of $fill_p$ (1.0 and 1.20). 

The mass function of the SB1 solution (Table \ref{solorbPshort}) can be written:
\begin{equation}
  \frac{m_s^3\,\sin^3{i}}{(m_p + m_s)^2}  =  \frac{P_{\rm orb}\,K_p^3}{2\,\pi\,G}\,(1 - e^2)^{3/2}
\end{equation}
which, given that $e = 0.0$, leads to
\begin{equation}
  \sin{i} = 6.38\,10^{-3}\,K_p\,\left(\frac{{\rm M}_{\odot}}{m_p}\,q\,(1+q)^2\right)^{1/3}
  \label{rel2}
\end{equation}
where $K_p$ is expressed in km\,s$^{-1}$. This relation is also displayed in Fig.\,\ref{massratio}, again for $m_p$ of 8 or 10\,M$_{\odot}$. The intersection between the curves corresponding to relations \ref{rel1} and \ref{rel2} indicate that the mass ratio of the short-period binary is likely close to 1.5, and that the orbital inclination is expected to be in the range between $22^{\circ}$ and $29^{\circ}$. Therefore, the secondary component of the short-period binary likely has a mass of $6^{+1.5}_{-1.2}$\,M$_{\odot}$, which would correspond to an B5 star with an uncertainty of one or two spectral substypes. 

We also revisited the analysis of the photometric lightcurve presented by \citet{naz20tess}, taking advantage of the better knowledge of the short-period binary. Adopting the improved ephemerides of the short-period binary (Table \ref{solorbPshort})\footnote{Note that an incorrect $T_0$ was mentioned in \citet{naz20tess}: the actual value to get Fig. A1 in that paper (with $\phi=0$ for the main eclipse) was 2\,458\,789.349. However, the new data now provide better ephemeris so that the old values can now be dismissed.}, we built an updated lightcurve consisting of 100 normal points computed as the mean magnitude per 0.01 phase bin. The dispersion of the data points within a phase bin is adopted as our estimate of the uncertainty. The {\tt Nightfall} binary star code \citep{Wichmann} is then used to fit the lightcurve. The model accounts for the presence of third light, in this case coming from the Be star, and for reflection effects between the stars. We fixed the mass ratio to 1.5 and the primary star effective temperature to 20\,900\,K. We explored a grid of values of the Roche-lobe filling factors between 1.00 and 1.20, and third light contributions ranging from 57\% to 62\%. There is a huge degeneracy between different models in the parameter space, but overall the lightcurve analysis confirms the orbital inclination to be in the 21$^{\circ}$ -- 30$^{\circ}$ range. The formally best solution for $q = 1.5$ (i.e. masses of 9 and 6\,$M_{\odot}$) is found for $fill_p = fill_s = 1.175$, polar radii of 8.7 and 7.5\,R$_{\odot}$, a third light contribution of 57\% (corresponding to V782\,Cas\,A), an orbital inclination of $21.5^{\circ}$ and a secondary mean temperature (accounting for the heating due to reflection effects) of 11\,710\,K. In this solution, the primary star contributes 35\% of the total light of V782\,Cas, whereas the secondary provides 8\% of the light. Those values agree well with those found from spectroscopy (see above). The lightcurve and its best-fit model are shown at the bottom of Fig. \ref{SB1HeI}. Because of the low inclination, the recorded variations are not true eclipses. Rather, the photometric changes are due to ellipsoidal variations. The unequal depth of the photometric minima is due to the difference in surface brightness distribution of the two stars, which is notably affected by reflection effects. Whilst the overall quality of the fit is rather good, some of the remaining deviations are likely due to the intrinsic variability of the Be star \citep{naz20tess}. The light ratio of about 4 between the two stars in V782\,Cas\,B agree well with their respective spectral types, but we note that the temperature of the secondary object is a little bit low for the spectral type of B5 mentioned above \citep{dri00}. However, because of the on-going interaction (both stars overfill their Roche lobes), the stars may present non-standard parameters.

\begin{figure}
  \begin{center}
      \includegraphics[width=8cm]{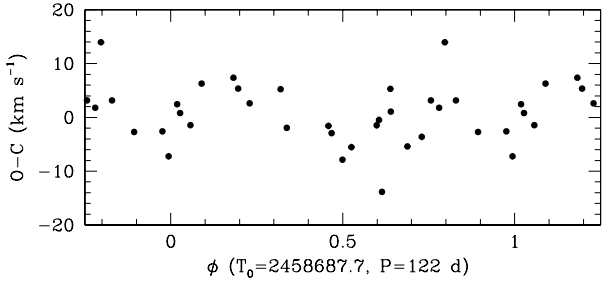}
  \end{center}
  \caption{O-C residuals of the SB1 solution of the primary component of the short-period binary in V782\,Cas folded with the ephemerides of the Be star. \label{OC}}
\end{figure}

\begin{figure*}
  \begin{center}
    \includegraphics[width=18cm,bb=30 510 575 700, clip]{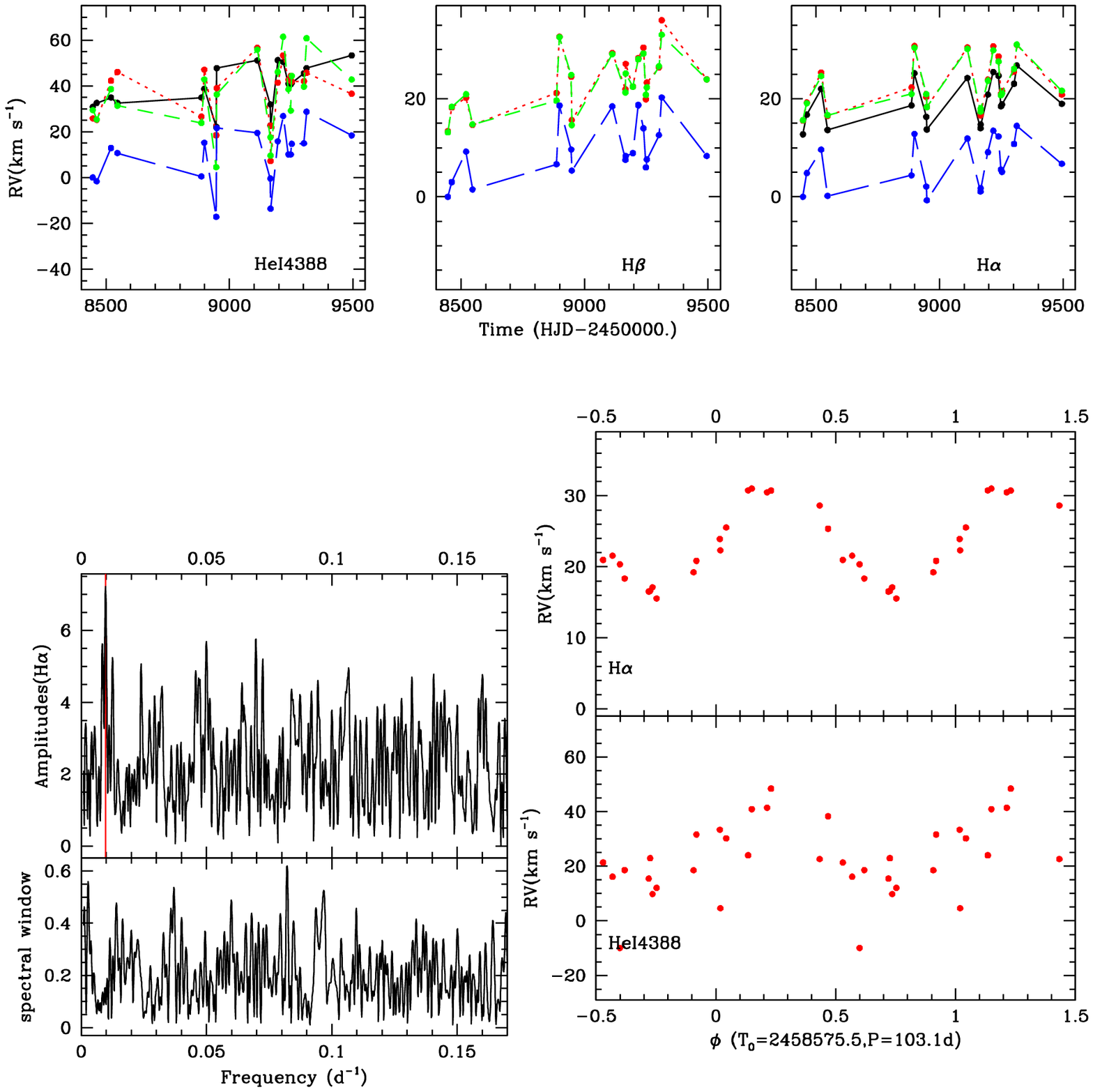}
  \end{center}
  \caption{Radial velocities measured for HD\,45995 on He\,{\sc i}\,4388\,\AA\ (left), H$\beta$ (middle) and H$\alpha$ (right) lines using the moment method (black points and solid line), the mirror method (red points and dotted line), the double Gaussian method (green points and dashed line), and the correlation method (blue points and long dashed line). Note that the mix of absorption and emission in H$\beta$ makes the first order moments unreliable hence the associated RVs are not shown. }
\label{hd45rv}
\end{figure*}

\begin{figure}
  \begin{center}
    \includegraphics[width=8cm,bb=30 150 285 425, clip]{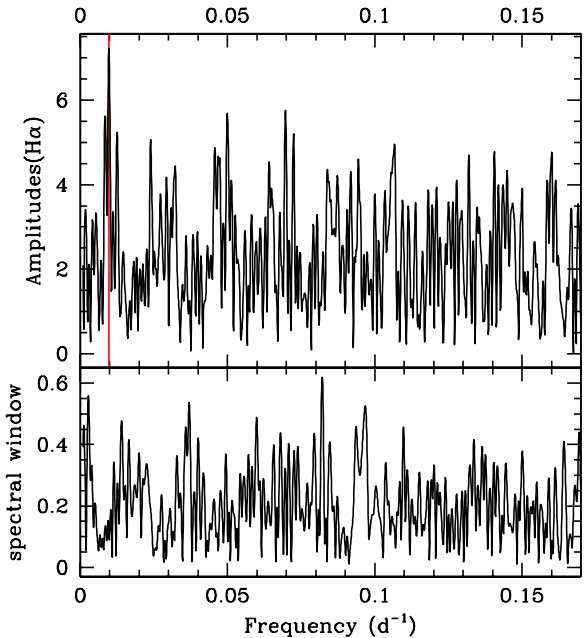}
  \end{center}
  \caption{Periodogram derived with the modified Fourier algorithm for H$\alpha$ velocities of HD\,45995 (measured with the mirror method), along with its spectral window. A vertical line indicates the proposed orbital period.}
\label{hd45fou}
\end{figure}

\begin{figure}
  \begin{center}
    \includegraphics[width=8cm,bb=290 150 575 500, clip]{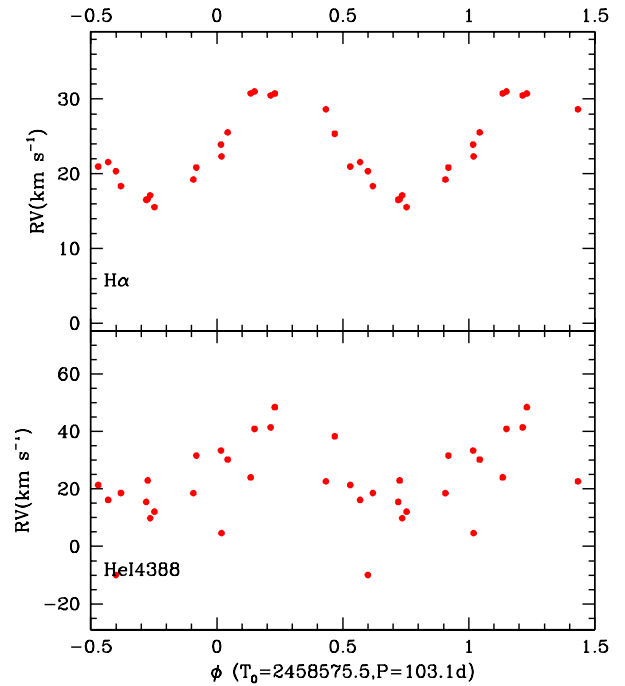}
  \end{center}
  \caption{Velocities of HD\,45995 measured with the mirror method for He\,{\sc i}\,4388\,\AA\ (bottom) and H$\alpha$ (top) folded with the best-fit ephemeris. }
\label{hd45phi}
\end{figure}

\subsubsection{V782\,Cas A, the long-period binary}
Turning to the long-period binary, we fitted the $RVs$ of the H$\alpha$ line by a sinusoid, resulting in an amplitude of 5.2$\pm$0.9\,\kms\ with $T_0=2\,458\,687.7\pm2.6$ (conjunction with the Be star in front). The velocity dispersion then decreases from 4.4\,\kms\ to 2.6\,\kms\ (see right panel of Fig. \ref{v782phi} - note the very good agreement between CARMENES and TIGRE data). The associated mass function is here 0.0018$\pm$0.0009\,$M_{\odot}$. Considering the Be star (spectral type B2.5III) to have a mass of 9\,$M_{\odot}$ and the system's inclination to be 60--90$^{\circ}$, the companion should then have a mass of 0.55--0.65$\pm$0.10\,$M_{\odot}$. The period is similar to that found for the $\phi$\,Per system \citep{mou15} but the amplitude is half in V782\,Cas. Hence the companion is less heavy, relative to the primary star and for a similar inclination. Note that V782\,Cas has not been studied by \citet{wan18}.

To check whether the short-period binary and the Be star are gravitationally bound with an orbital period of 122\,d, we folded the O-C residuals of the above SB1 solution of the short-period binary with the ephemerides of the Be star. The result is illustrated in Fig.\,\ref{OC}. Whilst we observe O-C values that range between $-14$ and $+14$\,km\,s$^{-1}$, there is no clear trend that would indicate a reflex motion with respect to the Be star. This suggests that the period of the Be star is not an orbital period around the short-period binary, but rather around another, unseen companion. 

\begin{figure*}
  \begin{center}
    \includegraphics[width=18cm,bb=30 510 575 700, clip]{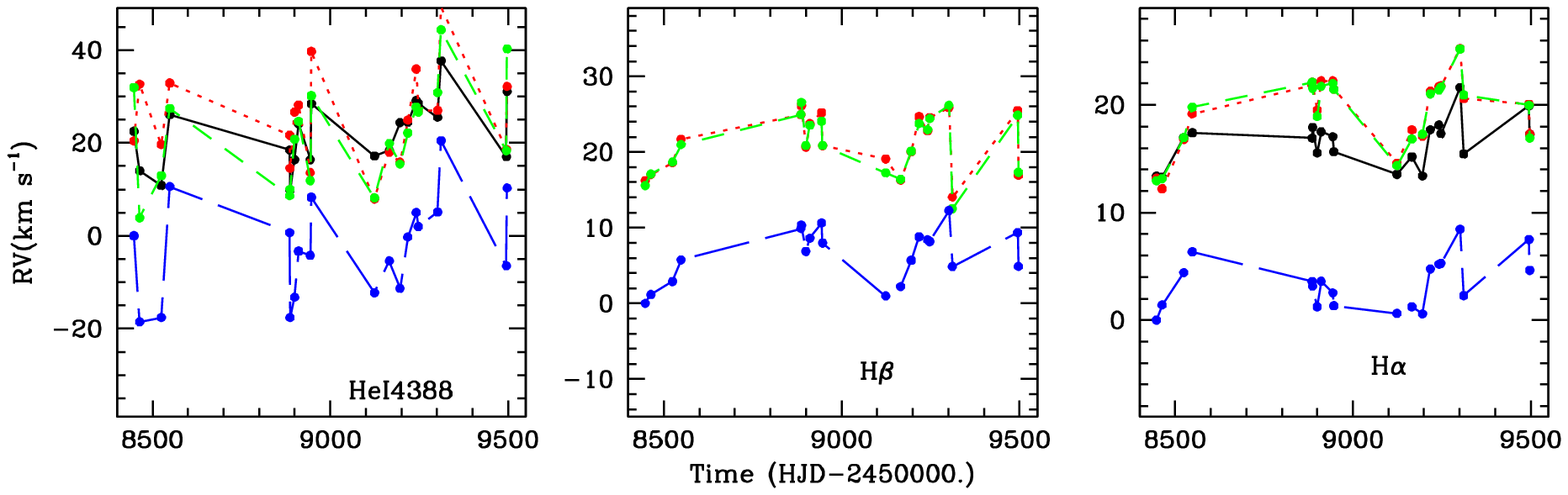}
  \end{center}
  \caption{Radial velocities measured for HD\,44458 on He\,{\sc i}\,4388\,\AA\ (left), H$\beta$ (middle) and H$\alpha$ (right) lines using the moment method (black points and solid line), the mirror method (red points and dotted line), the double Gaussian method (green points and dashed line), and the correlation method (blue points and long dashed line). Note that the mix of absorption and emission in H$\beta$ makes the first order moments unreliable hence they are not shown. }
\label{hd44rv}
\end{figure*}

\subsection{HD\,45995}

HD\,45995 was noted to have a companion with common proper motion, lying at 16.1\arcsec\ from the star, in \citet{abt84}. Its presence is confirmed in Gaia-DR3 although nothing is mentioned about it in \citet{mas99,rob07,mas09,mai10,hor20}. Large (50\,\kms) changes in radial velocity were reported by \citet{har87}. He pointed to a period shorter than 10\,d, preferentially 5.29\,d. Similar changes were also detected by \citet{gie86}, but they explained them in terms of non-radial pulsations of period 1.23\,d. Our analysis of \te\ photometry detected the presence of a frequency of 1.184\,d$^{-1}$ \citep{naz20tess}, of which the 1.23\,d period is a daily alias.

In the TIGRE data, variations of the radial velocities seem to be present but no obvious timescale is readily seen by eye (Fig. \ref{hd45rv}). We nevertheless calculated the periodogram associated to the H$\alpha$ velocities (derived with the mirror method). It reveals a peak for a period of 103.1$\pm$1.0\,d, with an amplitude triple of the mean level of the periodogram (Fig. \ref{hd45fou}). Folding the velocities with this period yields a very convincing result (Fig. \ref{hd45phi}). It must be noted that a clear sinusoidal modulation is observed for both H$\alpha$ and H$\beta$, but also for He\,{\sc i} lines although their RVs are noisier. A $\chi^2$ fit to H$\alpha$ velocities yields a best-fit sinusoid with parameters: $T_0=2\,458\,575.5\pm0.8$ (conjunction with Be primary in front) and $K=6.7\pm0.4$\,\kms. The H$\alpha$ velocity dispersion passes from an initial 5\,\kms\ to only 1.5\,\kms\ after that best-fit sinusoid is taken out.  The best-fit parameters yield a mass function of 0.0032$\pm$0.0006\,$M_{\odot}$. Considering the Be star (spectral type B2V) to have a mass of 10\,$M_{\odot}$ \citep{vie17} and the system's inclination to be 46.8$^{\circ}$, i.e. that of the disk \citep{fre05}, the companion should then have a mass of 1.0$\pm$0.1\,$M_{\odot}$. 

\subsection{HD\,44458}

This star is most probably associated with a 4\arcsec\ neighbouring star \citep{abt84,rob07}. In the latter reference, the companion is said to have a spectral type B8--A0\,V, with the Be star reported to be a Cepheid (!). No information about a closer companion is available, since no change in radial velocity was reported in literature. However, clear RV variations are seen in our data and they are relatively coherent from line to line (Fig. \ref{hd44rv}). Their amplitude is about 10\,\kms\ in H$\alpha$. While the velocities were analyzed with the different period search algorithms, no coherent picture emerges. Furthermore, the folding with the best-fit ``period'' found with the Fourier method (332$\pm$10\,d - suspiciously close to a year) is not totally convincing: the RVs are found to cover only two-third of the alleged orbit and to increase monotoneously in that interval. Therefore, while the presence of velocity variations seems ascertained and the star can be considered as a binary candidate, more observations are needed before assessing their recurrence. We may however already conclude that any recurrence timescale must be larger than the yearly observing campaign durations (100--300\,d).

\begin{table}
\centering
  \caption{Results from our velocity monitoring. For confirmed binaries, the period, amplitude, and zero-points of their $RV$ curves are provided; for binary candidates, approximate values for the temporal interval between the most distant exposures and for the maximum $RV$ shift are noted. The number of measurements $N$ is also listed, and the two possibilities for SAO\,49725 are quoted. $\gamma$ values depend on the method used: listed values are those found for velocities derived by the mirror method.}
\label{res}
\setlength{\tabcolsep}{3.3pt}
\begin{tabular}{lccccc}
  \hline\hline
\multicolumn{4}{l}{\it Confirmed binaries}\\
  Name & $N$ & $P$ & $K$     & $\gamma$ & $T_0$\\
       &     & (d) & (\kms ) & (\kms )  & (HJD-2\,450\,000)\\
  \hline
V782\,CasA &27+9&122.0$\pm$1.5 & 5.2$\pm$0.9 &  -5.6$\pm$0.2 & 8687.7$\pm$2.6 \\
HD\,45995  &19  &103.1$\pm$1.0 & 6.7$\pm$0.4 &  23.9$\pm$0.2 & 8575.5$\pm$0.8 \\
V558\,Lyr  &14  & 83.3$\pm$1.8 & 8.2$\pm$1.1 & -46.9$\pm$0.3 & 9056.2$\pm$1.4 \\
SAO\,49725 &11+1&26.11$\pm$0.08& 2.8$\pm$0.5 & -13.7$\pm$0.3 & 8712.1$\pm$0.3 \\
           &    &137.0$\pm$2.3 & 2.7$\pm$0.7 & -12.1$\pm$0.3 & 8780.0$\pm$1.5 \\
V2156\,Cyg &10+3&126.6$\pm$2.0 & 5.5$\pm$0.7 & -12.9$\pm$0.3 & 8812.0$\pm$1.2 \\
V810\,Cas  &10+3& 75.8$\pm$0.7 & 6.4$\pm$0.7 & -27.9$\pm$0.3 & 8753.0$\pm$2.0 \\
\hline
\multicolumn{4}{l}{\it Binary candidates}\\
  Name & $N$ &$\Delta(t)$ & $\Delta(RV)$  \\
       &     &(d) & (\kms ) \\
  \hline
HD\,44458  &20& 140 & 10\\
HD\,110432 & 6&  70 & 10\\
HD\,119682 & 6&  90 & 20\\
HD\,161103 & 5& 160 & 10\\
V771\,Sgr  & 7& 170 & 15\\
  \hline      
\end{tabular}
\end{table}

\subsection{Other northern stars}

\begin{figure*}
  \begin{center}
    \includegraphics[width=8.5cm]{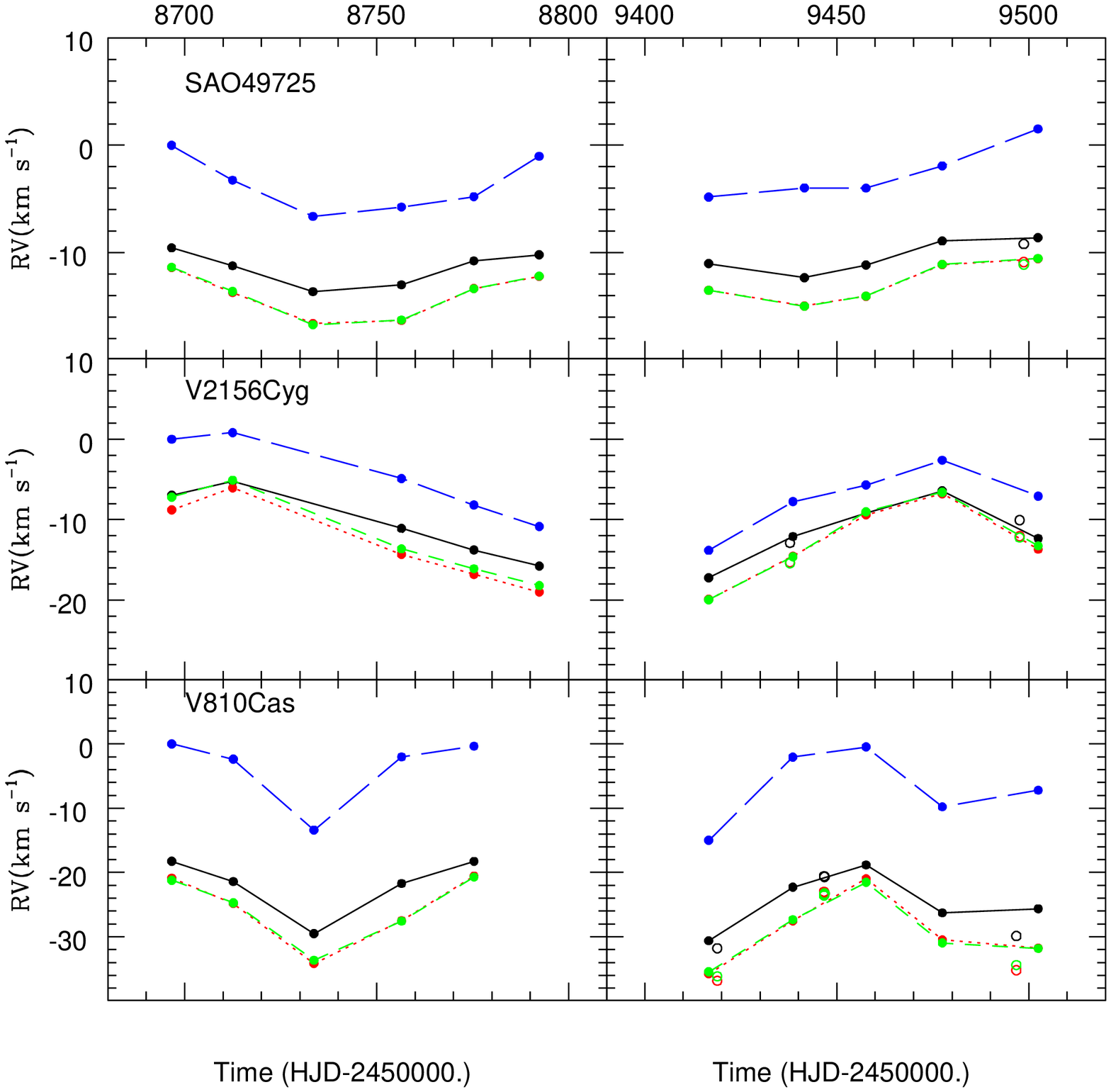}
    \includegraphics[width=8.5cm]{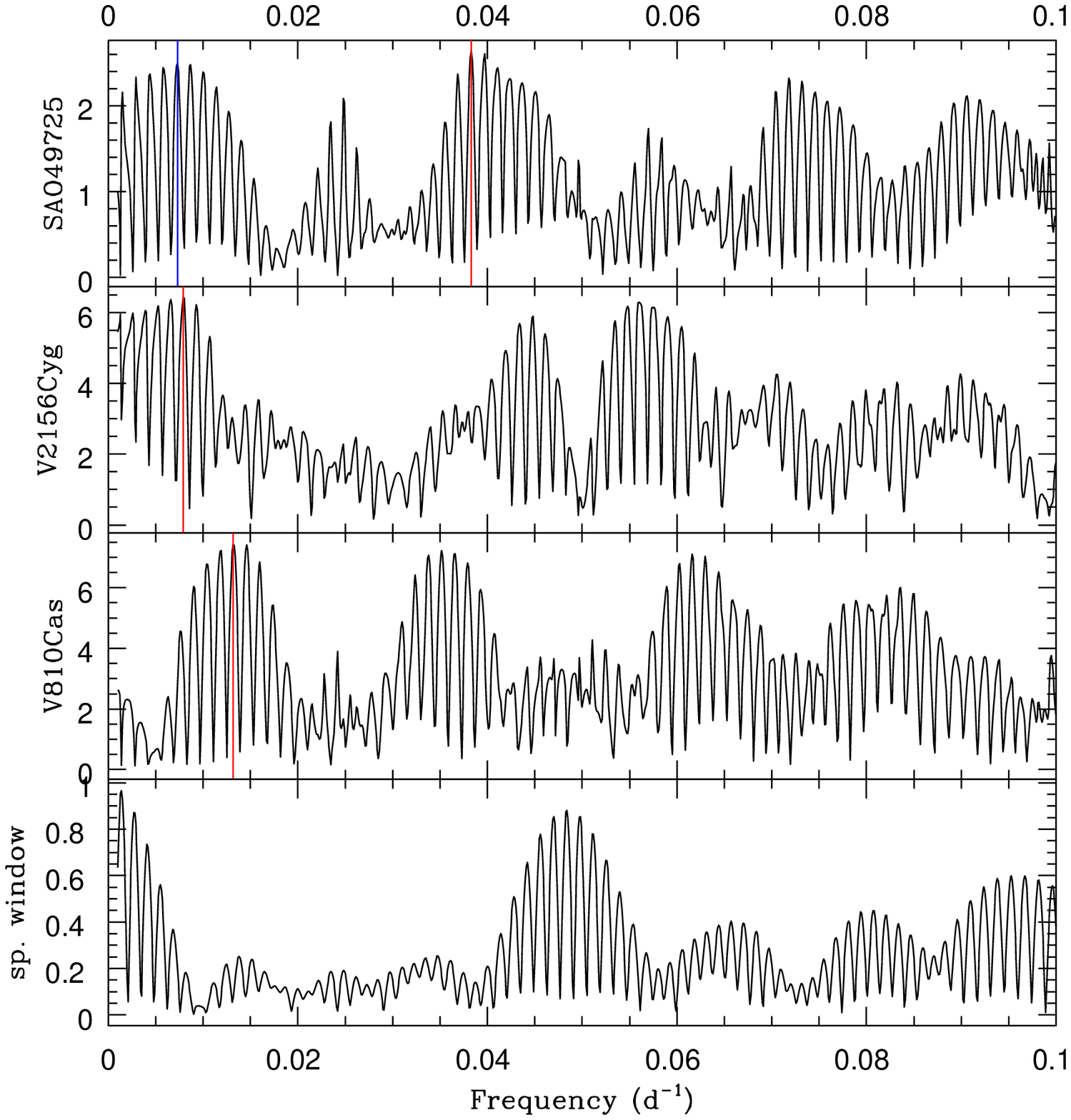}
  \end{center}
  \caption{{\it Left:} Velocities measured in 2019 (left) and 2021 (right) for SAO\,49725 (top), V2156\,Cyg (middle), and V810\,Cas (bottom). These velocities were derived from the H$\alpha$ lines using the moment method (black points and solid line), the mirror method (red points and dotted line), the double Gaussian method (green points and dashed line), and the correlation method (blue points and long dashed line). Filled points correspond to velocities derived from CARMENES observations, open symbols from TIGRE observations. {\it Right:} Periodograms derived from the CARMENES velocities using the mirror method. Vertical red and blue lines indicate the highest peaks chosen as orbital periods. }
\label{carme}
\end{figure*}

\begin{figure}
  \begin{center}
    \includegraphics[width=8.5cm]{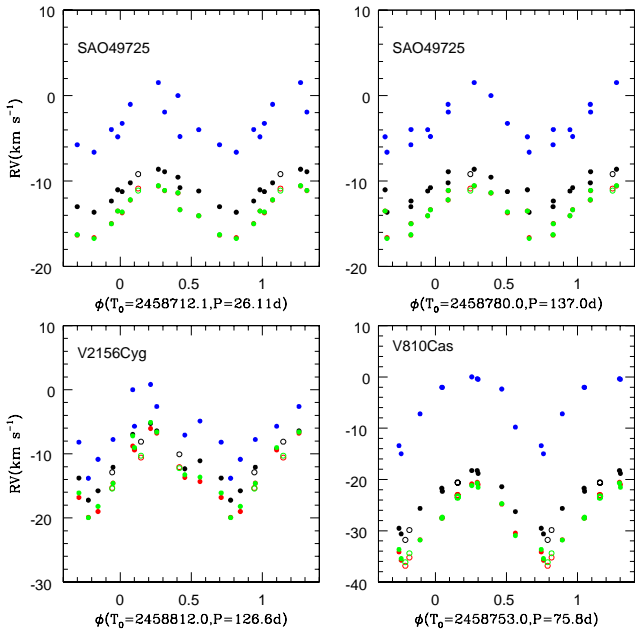}
  \end{center}
  \caption{Same velocities as in left panels of Fig. \ref{carme}, but phase-folded with the best-fit ephemeris.}
\label{carmeph}
\end{figure}

No optical companion was reported for SAO\,49725, V2156\,Cyg, and V810\,Cas \citep{abt84,mas99,rob07,mas09,mai10,hor20}. The CARMENES data of these three stars display clear RV variations, and these velocity changes agree for all measurements methods and for all considered lines (H, He, Fe). This is clearly reminiscent of binary motion. The velocity changes observed in 2019 were confirmed in the 2021 campaign (left panels of Fig. \ref{carme}). The number of velocity measurements ($\sim$10) is sufficient to attempt a period search (right panels of Fig. \ref{carme}). While some peaks can be detected, with an amplitude of about 2.5 times the mean level of the periodogram, numerous aliases are also present. For SAO\,49725, this even leads to two equally high peaks, corresponding to two plausible periods and even the inclusion of TIGRE data does not allow to distinguish them. Therefore the derived periods require confirmation. Nevertheless, as for the previous cases, we have derived the best-fit sinusoidal solutions using a $\chi^2$ fitting.

For SAO\,49725, the two preferred periods are 26.11$\pm$0.08\,d and 137.0$\pm$2.3\,d. The former value leads to an amplitude of 2.8$\pm$0.5\,\kms, which in turn yields a mass function of (5.9$\pm$3.2)$\times 10^{-5}$\,M$_{\odot}$, and $T_0=2\,458\,712.1\pm0.3$. For the latter values, one rather finds $K=2.7\pm0.7$\,\kms, $T_0=2\,458\,780.0\pm1.5$, and $f(m)=(2.8\pm2.2)\times 10^{-4}$\,M$_{\odot}$. In both cases, the velocity dispersions changes from 2 to 1\,\kms\ after the fitting. The star displays rather narrow lines, compared to other stars in this sample, suggesting that it is viewed under a lower inclination. Hence we enlarged the considered range of orbital inclinations to 30--90$^{\circ}$, and then find secondary masses of 0.22--0.45$\pm$0.05\,M$_{\odot}$ and 0.4--0.7$\pm$0.1\,M$_{\odot}$, respectively, considering a mass for the Be star intermediate between those of \gc\ and $\pi$\,Aqr (see below).

For V2156\,Cyg, \citet{cho17} reported RV variations from --20 to --5\,\kms\ on the Br11 hydrogen line at 1.68\,$\mu$m. While their dataset consisted of only three spectra, the presence of velocity variations is confirmed by our data. Our periodogram presents its strongest peak at 126.6$\pm$2.0\,d. The $\chi^2$ fit to H$\alpha$ velocities yields a best-fit sinusoid with parameters $T_0=2\,458\,812.0\pm1.2$ (conjunction with Be primary in front) and $K=5.5\pm0.7$\,\kms. The velocity dispersion passes from an initial 5\,\kms\ to only 2\,\kms\ after that best-fit sinusoid is taken out. These best-fit parameters yield a mass function of 0.0022$\pm$0.0008\,M$_{\odot}$. As the Be star has a spectral type of B1.52V, we may assume its mass to be about 11\,M$_{\odot}$ \citep{vie17}. With a system's inclination in the range 60 to 90$^{\circ}$, this leads to a companion mass of 0.7--0.8$\pm$0.1\,$M_{\odot}$. 

For V810\,Cas, the strongest peak is at 75.8$\pm$0.7\,d and the sinusoid fitting to the H$\alpha$ velocities yields $K=6.4\pm0.7$\,\kms\ and $T_0=2\,458\,753.0\pm2.0$, with a dispersion decreased from 5 to 1\,\kms. With a spectral type B1 (hence a mass of about 12.5\,M$_{\odot}$, \citealt{vie17}), we find $f(m)=0.0021\pm0.0007\,$M$_{\odot}$ and a secondary mass of 0.7--0.8$\pm$0.1\,$M_{\odot}$ for an inclination range of 60--90$^{\circ}$.

These solutions provide convincing phase plots (right panels of Fig. \ref{carmeph}). However, as mentioned above, all these orbital solutions should be considered as preliminary because of the aliases in the periodogram. More data should be collected to confirm them in the near future. It is however important to note that the few existing TIGRE measurements are fully consistent with the CARMENES data (Fig. \ref{carmeph}). 

\begin{figure*}
  \begin{center}
    \includegraphics[width=8.85cm,bb=30 170 575 715,clip]{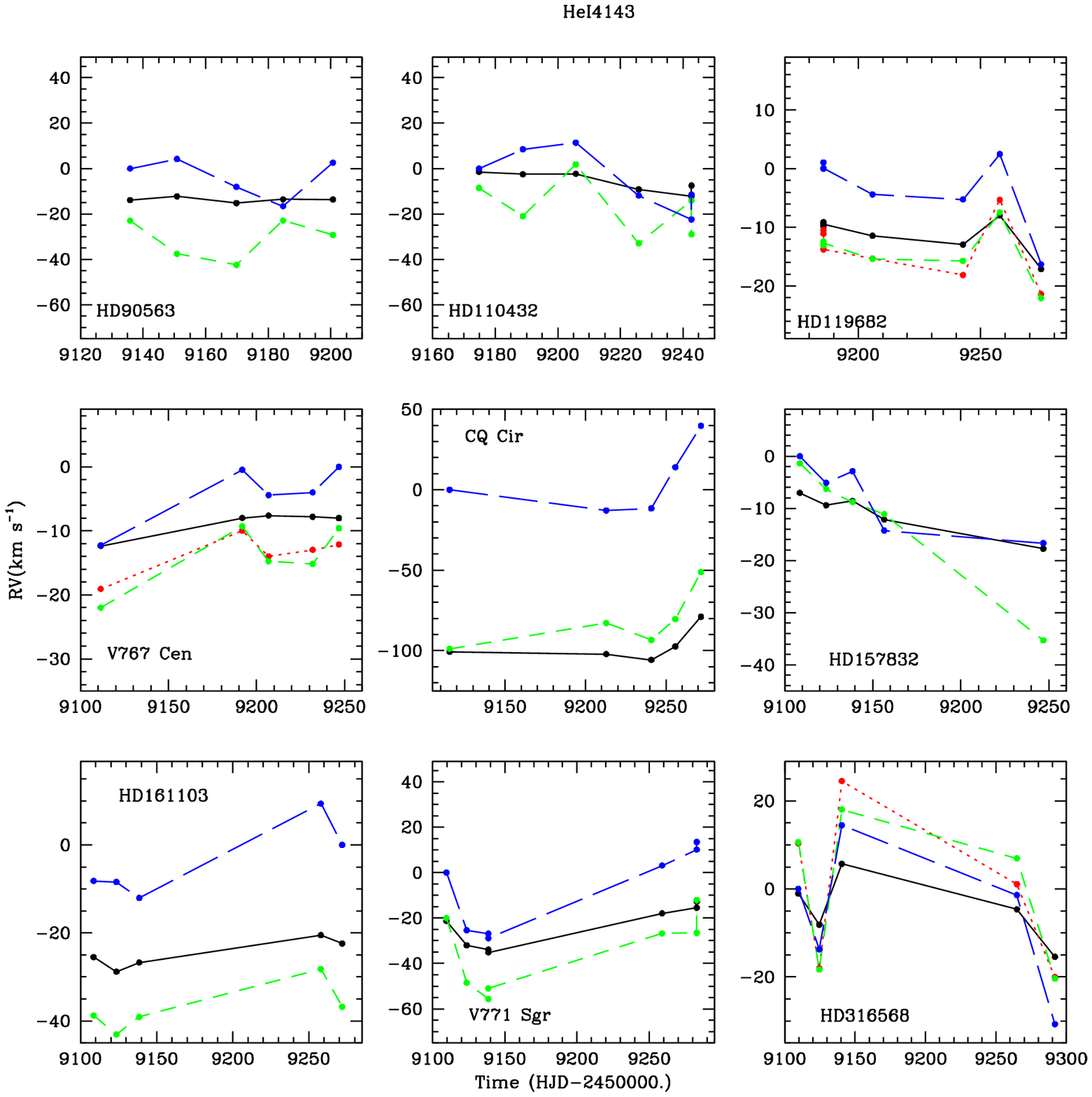}
    \includegraphics[width=8.85cm,bb=30 170 575 715,clip]{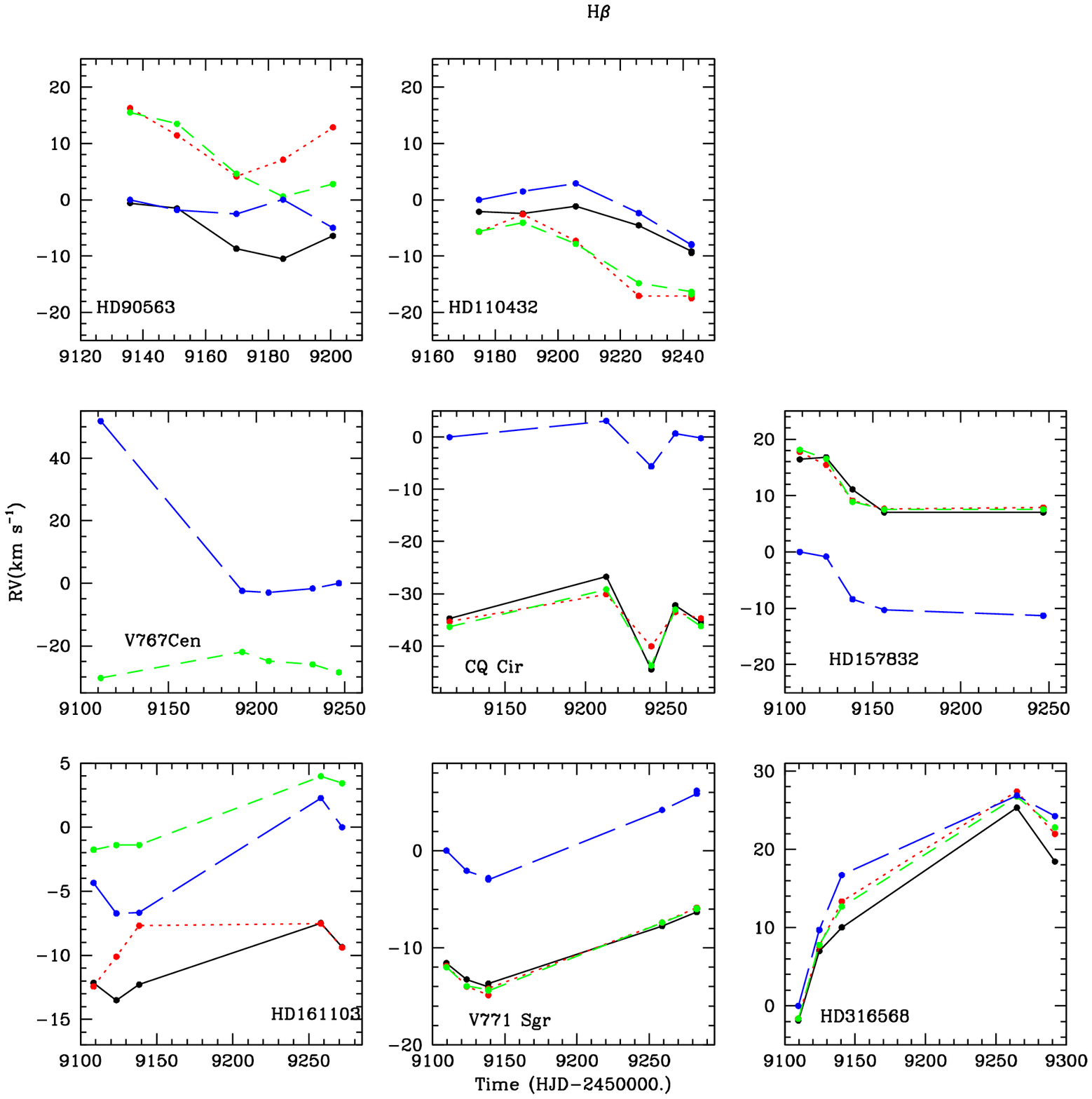}
  \end{center}
  \caption{Velocities of the He\,{\sc i}\,$\lambda$4143\,\AA\ line (left panels) and H$\beta$ line (right panels) recorded in the southern targets observed by UVES. Symbols are as in Fig. \ref{carme}. Four cases (HD\,110432, HD\,119682, HD\,161103, and V771\,Sgr) are kept as binary candidates.}
\label{uvesha}
\end{figure*}

\subsection{Southern stars}
The southern sample suffered from a lack of multiplicity investigation. Only V767\,Cen was mentioned to have a distant optical companion (a B8V star at 15\arcsec, \citealt{lin85}), while high-resolution interferometric data could not detect any companion to HD\,110432 \citep{ste13,san14}. In our UVES data, all stars display velocity changes (Fig. \ref{uvesha}). However, the line profiles also appear much more variable than for TIGRE and CARMENES targets, which could feign velocity changes unrelated to binarity. Besides, the detected trends are not always consistent amongst different lines, pointing to another probable cause than binarity. Therefore, we split the results in different categories.

Both HD\,157832 and HD\,316568 display prominent line profile variations, both in He\,{\sc i} absorptions and in the core of Balmer emissions. For example, one of the He\,{\sc i} profiles of HD\,157832 appears perfectly triangular while it is rounded at other dates, whereas the He\,{\sc i} profiles of HD\,316568 appear more or less skewed or with a flat bottom. This points towards pulsational activity, which is known to exist in HD\,157832 from photometric data \citep{naz20tess}. Therefore, we cannot attribute the observed velocity changes to binarity with any certainty.

Some stars rather display significant changes in overall disk activity. The first observation of V767\,Cen, taken three months before the four other ones, revealed a much reduced strength of the disk emissions. Discarding this outlier and focusing only on the last data, the velocities vary in a negligible way, showing no obvious binary signature. There are also clear changes in the disk emissions of HD\,119682, with an increased activity in the last observation. However, the line profiles of He\,{\sc i}\,$\lambda$4026,4143,4388\,\AA\ appear quite stable and their velocities display coherent trends. We therefore consider these observations as tentative evidence for binary motion.

For HD\,90563, the measured velocities do not agree well between methods and/or between lines. For CQ\,Cir, the mesurement methods agree for a given line but vary from line to line. This renders those stars unsuitable for entering a list of binary candidates, although more observations would be welcome to shed some light on the behaviour of these targets.

Coherent velocity changes between lines and between measurement methods are found for HD\,110432, HD\,161103, and V771\,Sgr, however. The relatively long variation timescales and the shallow amplitudes are reminiscent of what is seen in the known binaries (\gc, $\pi$\,Aqr, and the three new systems found in previous sections).

While evidence for binarity in the southern sample observed by UVES remains sparse so far, we find four cases (HD\,110432, HD\,119682, HD\,161103, and V771\,Sgr) showing promising hints. Therefore we grant those stars the status of binary candidates. As for CARMENES targets, further observations are required to confirm their multiplicity status and establish orbital solutions.

\section{Discussion}
Our monitoring campaign of \gc\ stars has revealed velocity changes in all stars, but not all of them could be securely attributed to binary motions. Amongst the 16 studied stars, a binary orbit could be well established for the Be star in three cases, a further three cases provided preliminary orbits, and five other stars can be considered as binary candidates (Table \ref{res}). Adding the two known binaries (\gc, $\pi$\,Aqr), this means that the velocity of 72\% of known \gc\ stars has been monitored. Eight out of 18 (half of them) display definitive signatures of binary motion, with periods and amplitudes of orbital solutions ascertained. When the binary candidates are added, nearly three quarters of the studied stars are or could be binaries. The binary incidence amongst \gc\ stars thus appears far from negligible. However, this fraction might only reflect the incidence of binaries amongst classical Be stars (see Section 1.1), rather than being a specific feature of \gc\ analogs.

In this context, it is important to examine the properties of orbits and companions of Be stars. To this aim, Table \ref{list} summarizes the properties of known Be binary systems. The top of the table provides the information available for the \gc\ stars while the bottom part lists cases with early-type (B0--B3) classical Be stars not known to belong to the \gc\ category. Because they represent a different evolutionary stage, we have excluded here the currently interacting cases in which the Be star actively {\it accretes} material from a red giant companion (e.g. AX\,Mon, V644\,Mon, CX\,Dra, V360\,Lac - \citealt{eli97,auf94,har15}). Furthermore, the table only presents systems with well-defined orbital solutions. This differs from the choices of \citet{bod20} who examined Be stars with spectral type B1.5e or earlier and split the ``systems'' depending on the companion's nature (post-MS/unknown). These authors considered some ambiguous indirect information as secure binarity evidence\footnote{In \citet{bod20}, HD\,161103 (=V3892\,Sgr) is categorized as having a ``known post-MS companion'' (their class i) while the provided references only offer a general discussion on the origin of the \gc\ phenomenon, with the magnetic and WD hypotheses highlighted: none of the provided references provide a direct, secure detection of a companion. Our velocity monitoring can only attribute it a ``binary candidate'' status.} while missing some secure binary cases such as the well-known $\phi$\,Per system. Focusing on systems with existing orbital solutions, as done here, is certainly more restrictive but, in our opinion, it helps avoiding any ambiguity.

\begin{table*}
\centering
  \caption{List of binaries with an early (B0--3) classical Be primary and an existing orbital solution, ordered by right ascension (R.A.). }
\label{list}
\setlength{\tabcolsep}{3.3pt}
\begin{tabular}{lccccccc}
  \hline\hline
  Name & $P$ & $e$ & Be Sp.type & $M$(Be) & $M_{\rm comp}$ & $i$ & Reference \\
       & (d) &     &            & (M$_{\odot}$) & (M$_{\odot}$) & ($^{\circ}$) & \\
  \hline
\multicolumn{8}{l}{\it \gc\ stars}\\
\gc        & 203.6 & 0 & B0IV    & 13 & 0.98$^*$       & 45      & \citet{nem12} \\
V782\,Cas  & 122.0 & 0 & B2.5III & 9  & 0.6--0.7$^*$   & 60--90  & this work     \\
HD\,45995  & 103.1 & 0 & B2V     & 10 & 1.0$\pm$0.1$^*$& 46.8    & this work     \\
V558\,Lyr  & 83.3  & 0 & B3V     & 8  & 0.7--0.8$^*$   & 60--90  & this work     \\
SAO\,49725 & 26.11 & 0 & B0.5III & 13 & 0.2--0.5$^*$   & 30--90  & this work\\
           & 137.0 & 0 &         &    & 0.4--0.7$^*$   & 30--90  & this work\\
V2156\,Cyg & 126.6 & 0 & B1.5V   & 11 & 0.7--0.8$^*$   & 60--90  & this work\\
$\pi$\,Aqr & 84.1  & 0 & B1V     & 15 & 2.4$\pm$0.5    & 70      & \citet{bjo02} \\
V810\,Cas  &  75.8 & 0 & B1      &12.5& 0.7--0.8$^*$   & 60--90  & this work\\
\hline
\multicolumn{8}{l}{\it Other Be stars}\\
$\phi$\,Per     & 126.7 & 0    & B1.5V    & 9.6      & 1.2$\pm$0.2   & 77.6 & \citet{mou15} \\
$\zeta$\,Tau    & 133.0 & 0    & B1IV     & 11       & 0.9--1.0$^*$  &60--90& \citet{ruz09} \\
HR\,2142        & 80.9  & 0    & B1.5IV-V & 10.5     & 0.7$^*$       & 85   & \citet{pet16} \\
LB-1            & 78.8  & 0    & B3V      & 7$\pm$2  &1.5$\pm$0.4$^*$& 39   & \citet{she20} \\
HD\,55606       & 93.8  & 0    & B2.5-3V  & 6.0--6.6 & 0.83--0.9     &75--85& \citet{cho18} \\
FY\,CMa         & 37.3  & 0    & B0.5IV   & 10--13   & 1.1--1.5      & $>$66& \citet{pet08} \\
$o$\,Pup        & 28.9  & 0    & B1IV     & 11--15   & 0.7--1.0$^+$  &      & \citet{kou12} \\
MX\,Pup         & 5.15  & 0.46 & B1.5III  & 15       & 0.6--6.6      & 5--50& \citet{car02} \\
$\chi$\,Oph     & 138.8 & 0.44 & B2V      & 10       & 1.7--2$^*$    &60--90& \citet{abt78} \\
HD\,161306      & 99.9  & 0    & B0       & 15       & 0.9$^+$       &      & \citet{kou14} \\
HR\,6819        & 40.3  & 0.04 & B2.5V    & 6        & 0.4--0.8$^*$  & 35   & \citet{gie20} \\
59\,Cyg         & 28.2  & 0.14 & B1.5V    & 6.3--9.4 & 0.6--0.9      &60--80& \citet{pet13} \\
60\,Cyg         & 146.6 & 0    & B1V      & 11.8     & 1.5--3.4      & $>$29& \citet{kou00} \\
  \hline      
\end{tabular}

{\scriptsize $^*$ indicates SB1 systems for which the secondary mass was estimated from the probable values of primary mass, mass function, and inclination. $^+$ indicates SB2 systems for which the secondary mass was estimated from the mass ratio (derived from velocity amplitudes) and a probable value of the primary mass. When unconstrained, inclination is taken to be 60--90$^{\circ}$. Note that, for \gc, the results of \citet{smi12} and \citet{nem12} are similar - only the latter is quoted here. Velocity shifts have also been recorded for the secondary in HD\,194335, although a full orbital coverage has not yet been acquired \citep{wan21} hence we do not yet add it to this Table. } 
\end{table*}

Table \ref{list} clearly reveals some common features amongst Be binaries: periods are generally long (a few months\footnote{Note that the period of \gc\ is at the long period extreme of the period distribution in this Table, but this system went through intensive radial velocity searches by many authors. It thus may be expected that more extensive searches in other stars not yet known to be in binaries could ultimately lead to other long period cases.}) and the eccentricities small. This is valid for the \gc\ stars as for the other Be stars. Regarding the secondary mass, seven early Be have companion with masses below one solar mass, two with masses of 1--1.5\,$M_{\odot}$, and two with masses larger than 1.5\,$M_{\odot}$. For \gc\ stars, the respective numbers in these categories are 5, 2, and one. There is thus a prevalence of low-mass ($<1.5\,M_{\odot}$) companions in the known systems.

What does it imply on the nature of the companions? Considering first non-degenerate companions, such masses are compatible with late-type main-sequence stars (F or later) as well as with stripped Helium stars. However, it would be difficult to explain a strict preference towards late-type main-sequence stars when O and B-type systems rather display a preference towards early-type companions, i.e. $q$ ratios tend to be just less than 1 \citep[e.g.][]{san12}. Also, such low-mass stars should not display the UV signatures detected for some of the systems listed at the bottom of Table \ref{list}; rather, this type of signature is expected for hot stripped stars \citep{wan18,wan21}. Furthermore, in observations, there may be hints of a slight trend of increasing secondary masses with larger primary masses, a trend which is expected in stripped-star evolutionary models \citep{sha21}. Finally, the range of observed masses is compatible with that expected for stripped stars \citep{sha14,wan21}.

Secondly, compact objects are also possible in principle, but a single type cannot explain all companions. Indeed, neutron stars tend to have masses of 1--2.5\,$M_{\odot}$ \citep{als18} and can thus only explain the most massive (and rarer) companions. In contrast, white dwarfs generally have masses $<1\,M_{\odot}$ \citep{nal05,kil18} and, even if a few cases with masses just below Chandrasekhar's limit are known (e.g. \citealt{cai21}), they certainly cannot explain the most massive companions ($>1.5\,M_{\odot}$). 

We are thus left with the stripped stars as the most probable nature for the companions of Be stars, including \gc\ stars, if a single type of companions were to explain all binaries. Which impact does this conclusion have on the nature of the \gc\ phenomenon? Probably the scenario with a NS in propeller stage can now be discarded, as the low masses of most detected companions of \gc\ stars are not compatible with a NS nature \citep[see also][for other arguments]{smi17}. Furthermore, the fact that \gc\ stars do not appear significantly different from other Be stars - as much as the small number statistics seems to suggest - may not be favorable to scenarios in which the companion is responsible of the \gc\ phenomenon. Indeed, if the vast majority of Be stars lie in binaries and if the companion is responsible for the \gc\ phenomenon, then the vast majority of Be systems should display the same X-ray characteristics. This is not indicated by X-ray surveys of Be stars \citep{naz18}. This is also not suggested by the quite strict upper limits $\log(L_{\rm X}/L_{\rm BOL})<-7.3$ (or lower) on X-ray luminosities derived by \citet{ber96} for eight stars at the bottom of Table \ref{list}. Certainly more research is now needed to understand the \gc\ phenomenon. For example, to maintain a binary scenario would require some specific difference for the \gc\ systems (not period/separation, eccentricity, or secondary mass - but maybe peculiar disk/orbit  configurations or some other particular property?).

A final piece of information is that the peculiar tangential velocities of stars studied in this paper, derived from GAIA-eDR3, are all below 20\,\kms. Therefore, none of them seems to be a runaway although \citet{ber01} had proposed such a status for V782\,Cas. 

\subsection{Higher multiplicity}
Spectroscopy and photometry reveal the presence of two periods (2.5\,d and 122\,d) in V782\,Cas data, hinting at a high multiplicity. Our analysis shows that the short-period system is composed of two B-stars but there is no evidence that this binary is orbited on the longer timescale by the Be star, suggesting the system to be quadruple. 
           
High multiplicity may actually be common for massive O-stars \citep{san14}. Moreover, this is not the first time that high multiplicity is reported for Be stars. Several examples of visual companions have been reported, even for \gc\ itself\footnote{\citet{mam17} proposed this star to be triple and associated to the quadruple HD\,5408}. Additional companions have been proposed following spectroscopic analyses. In their study of $o$\,Cas (a system with $P\sim1032$\,d), \citet{kou10} concluded that, to explain its too large mass, the companion to the Be primary actually may be a short-period ($P\sim1.26$\,d) A+A binary. For CW\,Cep and 66\,Oph, periods of 2.7\,d and 10.78\,d were reported by \citet{sti92} and \citet{ste04}, respectively. The lack of motion for the H$\alpha$ line of these two Be stars suggested to \citet{riv20} and \citet{ste04}, respectively, that the Be star was a distant companion to an inner, short-period massive binary composed of two B-type stars (early-type for CW\,Cep and late-type for 66\,Oph). Note that both systems also have visual companions which are even further away. From a  similar observation, \citet{riv20} arrived at the same conclusion for LB-1 and for HR\,6819. However, these results ignited some lively discussions. For example, \citet{joh19} rather interpreted the lack of velocity shifts in H$\alpha$ as the sign of its arising in a circumbinary envelope: these authors considered CW\,Cep to be a binary, not a triple. In addition, \citet{she20} and \citet{gie20} reported reflex binary motions for H$\alpha$ in LB-1 and HR\,6819, respectively, discarding the ``outer Be in a triple system'' interpretation in favor of a more classical Be binary scenario. 

There are however less disputed cases of triple-star systems involving a Be star. \citet{mir13} reported on a long-term spectroscopic monitoring of $\delta$\,Sco. It comprises a Be star and a B1--3 companion in a highly eccentric orbit of period of 10.8\,yrs, but longer-term velocity variations indicate the presence of a third star in the system. In parallel, the spectrum of HD\,93683 indicates not only short-term binary motion linked to an inner O9V+B0V binary with $P\sim18$\,d, but also long-term velocity shifts in the H$\alpha$ line, suggesting the presence of an outer Be star with a period of about 400\,d \citep{bod20}. Interferometric and spectroscopic data further indicate that $\nu$\,Gem comprises an inner B+B binary with period 54\,d orbited in 19\,yr by an outer Be star \citep{gar21,kle21}.

The landscape of multiple systems comprising a Be star appears quite varied and, due to small number statistics, it is difficult to assess how similar/different V782\,Cas is, compared to previously known cases. 

\section{Conclusion}
We used high-resolution spectroscopy (TIGRE, CARMENES, UVES) to assess the multiplicity of a large sample of \gc\ stars. The remaining objects are either known to be binaries - two cases - or too faint for a thorough study with current facilities. An SB1 orbital solution could be established for six \gc\ stars, with one of them actually being a rare candidate of quadruple system. A further five stars show velocity shifts reminiscent of orbital motion, although more data are needed to fully establish the orbit. Combining these results with previous ones (\gc\ and $\pi$\,Aqr being known binaries) implies that at least a half of the studied \gc\ stars have companions and a further quarter are promising binary candidates. 

The derived parameters are in line with those found in other Be binaries: long periods, small eccentricities, and low-mass companions. This suggests that a single type of compact objects are probably not a good way to explain the \gc\ phenomenon as a whole. Also, the particularly bright and hard thermal X-ray emission remains the sole difference between \gc\ stars and other Be stars: the simple presence of companions cannot explain this peculiarity. 

\section*{Acknowledgements}
The authors acknowledge interesting discussions with Pr Anatoly Miroshnichenko and thank Pr Andrei Tokovinin for his constructive report. Y.N. and G.R. acknowledge support from the Fonds National de la Recherche Scientifique (Belgium), the European Space Agency (ESA) and the Belgian Federal Science Policy Office (BELSPO) in the framework of the PRODEX Programme (contracts linked to XMM-Newton and Gaia). ADS and CDS were used for preparing this document. TIGRE is a collaboration of the Hamburger Sternwarte, the Universities of Hamburg, Guanajuato and Li\`ege. This project is based on observations collected at CAHA (proposal H19-3.5-051+21B-3.5-051) operated jointly by junta de Andalucia and Consejo Superiod de Investigationes Cientificas (CSIC). Because it relies on OPTICON transnational access (program ID 19B/001), this project has received funding from the European Union's Horizon 2020 research and innovation programme under grant agreements No 730890 +101004719. This material reflects only the authors views and the Commission is not liable for any use that may be made of the information contained therein.

\section*{Data availability}
The ESO and Calar Alto (CARMENES) data used in this article are available in their respective public archives. The TIGRE data are available upon reasonable request.

\appendix
\section{Supplementary Data}
This Appendix provides the list of targets as well as the radial velocities obtained for each star. For V782\,Cas\,B, the velocities correspond to those found from the disentangling. For other targets, the velocities derived by the four methods are listed - note that for correlation, one spectrum per dataset serves as reference and has a zero velocity (there is one dataset per facility). For TIGRE and CARMENES data, the provided velocities were measured on the H$\alpha$ line while for UVES data, velocities of the H$\beta$ line are reported as the H$\alpha$ line may be subject to saturation in some cases. The two exceptions are HD\,119682 and V767\,Cen for which H$\beta$ has a complex profile mixing absorption and emission: He\,{\sc i}\,$\lambda$4143\AA\ velocities are provided instead.

\begin{table}
  \caption{List of targets examined in this paper \label{liststar}}
\begin{center}
  \begin{tabular}{lccc}
    \hline
    Name & Alt. Name & sp. type & $V$ (mag) \\
    \hline
V782\,Cas 	&HD\,12882 	&B2.5III  &7.62 	\\
HD\,44458 	&FR\,CMa 	&B1Vpe    &5.55 	\\
HD\,45995 	& 	        &B2Vnne   &6.14 	\\
HD\,90563 	& 	        &B2Ve     &9.86 	\\
HD\,110432 	&BZ\,Cru 	&B0.5IVpe &5.31 	\\
HD\,119682 	&        	&B0Ve     &7.90 	\\
V767\,Cen 	&HD\,120991 	&B2Ve     &6.10 	\\
CQ\,Cir 	&HD\,130437 	&B1Ve     &10.0 	\\
HD\,157832 	&V750\,Ara 	&B2Vne    &6.66 	\\
HD\,161103 	&V3892\,Sgr 	&Oe       &9.13 	\\
V771\,Sgr 	&HD\,162718 	&B3/5ne   &9.16 	\\
HD\,316568 	& 	        &B2IVpe   &9.66 	\\
V558\,Lyr	&HD\,183362 	&B3Ve     &6.34 	\\
SAO\,49725	&BD+47\,3129 	&B0.5IIIe &9.27 	\\
V2156\,Cyg 	&BD+43\,3913 	&B1.5Vnnpe&8.91 	\\
V810\,Cas 	&HD\,220058 	&B1npe    &8.59 	\\
\hline
  \end{tabular}
  \end{center}
\end{table}

\begin{table}
  \caption{Radial velocities measured on TIGRE data for V782\,Cas\,B using the disentangling method (see text for details). \label{rv782b}}
\begin{center}
  \begin{tabular}{lc}
    \hline
HJD-2\,450\,000. & $RV$ (\kms)\\
    \hline
8517.575 &   34.03   \\
8660.940 &   21.56   \\
8662.937 &   64.32   \\
8690.964 &   37.86   \\
8709.913 &  -48.16   \\
8728.878 &    5.26   \\
8744.749 &  -86.37   \\
8760.725 &   29.16   \\
8771.666 &   -8.74   \\
8788.816 &   53.08   \\
8806.733 &   13.08   \\
8820.674 &  -21.23   \\
8833.632 &   38.96   \\
8848.646 &   30.35   \\
8865.642 &  -68.31   \\
8887.597 &  -50.39   \\
9023.952 &  -65.95   \\
9055.935 &  -46.54   \\
9117.801 &   43.62   \\
9142.720 &   25.54   \\
9162.687 &    6.23   \\
9182.662 &  -14.19   \\
9203.670 &   38.11   \\
9236.591 &   -8.35   \\
9250.620 &   -9.60   \\
9418.947 &  -12.55   \\
9497.737 &   33.98   \\
\hline
  \end{tabular}
  \end{center}
\end{table}

\begin{table}
  \caption{Radial velocities (in \kms) measured on the H$\alpha$ line of V782\,Cas\,A using four methods (first-order moment, mirror, double-Gaussian, correlation - see text for details). \label{rv782a}}
\begin{center}
  \begin{tabular}{lcccc}
    \hline
HJD-2\,450\,000. & $M1$ & mirror & 2-Gaussian & correlation \\
    \hline
\multicolumn{5}{l}{\it TIGRE}\\
8517.575 & -5.3 & -8.3 & -6.8 & 10.0 \\ 
8660.940 & -10.1 & -11.4 & -9.7 &  4.0 \\ 
8662.937 & -10.9 & -10.7 & -9.5 &  3.1 \\ 
8690.964 & -6.0 & -6.1 & -5.0 &  8.0 \\ 
8709.913 & -3.4 & -1.9 & -1.0 &  9.8 \\ 
8728.878 & -2.9 & -1.5 & -0.3 & 13.3 \\ 
8744.749 & -9.5 & -8.7 & -5.8 &  3.4 \\ 
8760.725 & -8.9 & -11.6 & -9.3 &  7.9 \\ 
8771.666 & -13.3 & -13.1 & -11.3 &  0.0 \\ 
8788.816 & -9.4 & -9.9 & -8.5 &  5.2 \\ 
8806.733 & -7.1 & -7.3 & -6.3 &  5.5 \\ 
8820.674 & -5.4 & -4.8 & -3.8 &  7.9 \\ 
8833.632 & -1.4 &  0.6 &  1.2 & 12.1 \\ 
8848.646 & -0.8 &  0.4 &  0.7 & 15.1 \\ 
8865.642 & -6.9 & -6.2 & -3.9 &  5.6 \\ 
8887.597 & -11.2 & -12.1 & -10.4 &  2.1 \\ 
9023.952 & -12.4 & -12.5 & -10.6 &  2.0 \\ 
9055.935 & -5.5 & -5.3 & -4.2 &  8.4 \\ 
9117.801 & -2.8 & -5.0 & -0.9 & 14.2 \\ 
9142.720 & -7.7 & -8.8 & -7.7 &  5.9 \\ 
9162.687 & -4.7 & -6.1 & -4.9 &  8.8 \\ 
9182.662 & -0.7 &  0.9 &  2.0 & 13.7 \\ 
9203.670 &  0.1 &  2.7 &  3.2 & 13.0 \\ 
9236.591 & -0.5 & -3.4 & -0.8 & 15.0 \\ 
9250.620 & -9.6 & -10.8 & -10.0 &  4.6 \\ 
9418.947 & -5.2 & -3.3 & -2.9 &  8.6 \\ 
9497.737 & -8.7 & -8.3 & -7.1 &  5.2 \\
\hline
\multicolumn{5}{l}{\it CARMENES}\\
8696.651 & -7.5 & -6.6 & -5.6 &  0.0 \\ 
8712.654 & -2.5 & -0.4 &  0.2 &  3.7 \\ 
8733.675 &  0.2 & -0.8 & -0.2 &  9.9 \\ 
8756.488 & -4.0 & -7.4 & -4.7 &  4.4 \\ 
8775.538 & -11.1 & -11.9 & -10.4 & -5.3 \\ 
9416.663 & -3.2 & -2.2 & -2.2 &  4.1 \\ 
9457.644 & -1.3 & -2.3 & -1.8 &  7.4 \\ 
9477.489 & -0.8 & -2.0 & -0.2 &  7.3 \\ 
9502.462 & -6.1 & -6.0 & -5.4 &  0.2 \\ 
\hline
  \end{tabular}

{\scriptsize The line was integrated from -550 to 550\,\kms\ for the moment method. Wings with normalized fluxes between 1.4 and 2.2 were used for the mirror method. The double Gaussians had center velocities $\pm a = \pm$250\,\kms.}
  \end{center}
\end{table}

\begin{table}
  \caption{Same as Table \ref{rv782a} but for HD\,44458. \label{rv44458}}
\begin{center}
  \begin{tabular}{lcccc}
    \hline
HJD-2\,450\,000. & $M1$ & mirror & 2-Gaussian & correlation \\
    \hline
8447.873 & 13.4 & 13.2 & 13.0 &  0.0 \\ 
8463.801 & 13.3 & 12.2 & 13.2 &  1.4 \\ 
8524.664 & 17.0 & 16.8 & 17.0 &  4.4 \\ 
8548.618 & 17.4 & 19.2 & 19.8 &  6.3 \\ 
8885.714 & 16.9 & 21.9 & 22.1 &  3.6 \\ 
8886.624 & 17.9 & 22.1 & 21.6 &  3.2 \\ 
8899.632 & 15.5 & 19.5 & 18.9 &  1.2 \\ 
8910.661 & 17.5 & 22.2 & 21.7 &  3.6 \\ 
8943.622 & 17.1 & 22.3 & 22.0 &  2.5 \\ 
8946.585 & 15.7 & 21.5 & 21.5 &  1.3 \\ 
9123.953 & 13.6 & 14.6 & 14.4 &  0.6 \\ 
9165.846 & 15.2 & 17.7 & 16.8 &  1.2 \\ 
9195.791 & 13.4 & 17.1 & 17.3 &  0.6 \\ 
9217.819 & 17.7 & 21.3 & 21.1 &  4.8 \\ 
9241.688 & 18.1 & 21.7 & 21.4 &  5.2 \\ 
9247.647 & 17.3 & 21.8 & 21.7 &  5.3 \\ 
9301.588 & 21.6 & 25.3 & 25.2 &  8.5 \\ 
9311.583 & 15.5 & 20.6 & 20.9 &  2.3 \\ 
9494.977 & 19.9 & 20.1 & 20.0 &  7.5 \\ 
9496.997 & 17.3 & 17.3 & 16.9 &  4.6 \\ 
\hline
  \end{tabular}

{\scriptsize The line was integrated from -400 to 400\,\kms\ for the moment method. Wings with normalized fluxes between 1.6 and 2.8 were used for the mirror method. The double Gaussians had center velocities $\pm a = \pm$200\,\kms.}
  \end{center}
\end{table}

\begin{table}
  \caption{Same as Table \ref{rv782a} but for HD\,45995. \label{rv45995}}
\begin{center}
  \begin{tabular}{lcccc}
    \hline
HJD-2\,450\,000. & $M1$ & mirror & 2-Gaussian & correlation \\
    \hline
8446.917 & 12.7 & 15.5 & 15.6 &  0.0 \\ 
8462.819 & 16.7 & 19.2 & 19.1 &  4.8 \\ 
8520.683 & 22.0 & 25.4 & 24.6 &  9.6 \\ 
8546.628 & 13.6 & 16.5 & 16.7 &  0.2 \\ 
8886.720 & 18.6 & 22.3 & 21.0 &  4.4 \\ 
8898.659 & 25.2 & 30.8 & 30.4 & 12.8 \\ 
8946.593 & 16.3 & 20.4 & 20.9 &  2.1 \\ 
8948.628 & 13.7 & 18.4 & 18.3 & -0.7 \\ 
9112.990 & 24.3 & 30.5 & 30.2 & 11.9 \\ 
9165.816 & 14.0 & 16.6 & 17.4 &  1.1 \\ 
9166.823 & 14.7 & 17.1 & 18.0 &  1.8 \\ 
9195.801 & 20.8 & 23.9 & 23.7 &  9.1 \\ 
9217.791 & 25.5 & 30.7 & 29.9 & 13.5 \\ 
9238.713 & 24.7 & 28.6 & 27.6 & 12.3 \\ 
9248.649 & 18.5 & 21.0 & 20.7 &  5.6 \\ 
9252.681 & 18.9 & 21.6 & 21.1 &  5.1 \\ 
9301.606 & 23.1 & 25.5 & 26.1 & 10.8 \\ 
9312.574 & 26.8 & 31.0 & 31.1 & 14.5 \\ 
9494.993 & 19.0 & 20.8 & 21.6 &  6.7 \\ 
\hline
  \end{tabular}

{\scriptsize The line was integrated from -500 to 500\,\kms\ for the moment method. Wings with normalized fluxes between 1.4 and 2.2 were used for the mirror method. The double Gaussians had center velocities $\pm a = \pm$220\,\kms.}
  \end{center}
\end{table}

\begin{table}
  \caption{Radial velocities (in \kms) measured on the H$\beta$ line of HD\,90563 using four methods (first-order moment, mirror, double-Gaussian, correlation - see text for details). \label{rv90563}}
\begin{center}
  \begin{tabular}{lcccc}
    \hline
HJD-2\,450\,000. & $M1$ & mirror & 2-Gaussian & correlation \\
    \hline
9135.849 & -0.6 & 16.3 & 15.5 &  0.0 \\ 
9150.832 & -1.5 & 11.4 & 13.5 & -1.8 \\ 
9169.810 & -8.7 &  4.1 &  4.6 & -2.5 \\ 
9184.718 & -10.5 &  7.1 &  0.6 &  0.0 \\ 
9200.682 & -6.4 & 12.9 &  2.8 & -5.0 \\ 
\hline
  \end{tabular}

{\scriptsize The line was integrated from -400 to 400\,\kms\ for the moment method. Wings with normalized fluxes between 1.1 and 1.3 were used for the mirror method. The double Gaussians had center velocities $\pm a = \pm$200\,\kms.}
  \end{center}
\end{table}

\begin{table}
  \caption{Same as Table \ref{rv90563} but for HD\,110432. \label{rv110432}}
\begin{center}
  \begin{tabular}{lcccc}
    \hline
HJD-2\,450\,000. & $M1$ & mirror & 2-Gaussian & correlation \\
    \hline
9174.824 & -2.1 & -5.7 & -5.6 &  0.0 \\ 
9188.827 & -2.4 & -2.6 & -4.1 &  1.5 \\ 
9205.744 & -1.2 & -7.3 & -7.8 &  2.9 \\ 
9225.856 & -4.5 & -17.1 & -14.8 & -2.4 \\ 
9242.710 & -9.4 & -17.5 & -16.8 & -7.9 \\ 
\hline
  \end{tabular}

{\scriptsize The line was integrated from -400 to 400\,\kms\ for the moment method. Wings with normalized fluxes between 1.2 and 1.6 were used for the mirror method. The double Gaussians had center velocities $\pm a = \pm$200\,\kms.}
  \end{center}
\end{table}

\begin{table}
  \caption{Radial velocities (in \kms) measured on the He\,{\sc i}\,$\lambda$4143\AA\ line of HD\,119682 using four methods (first-order moment, mirror, double-Gaussian, correlation - see text for details). \label{rv119682}}
\begin{center}
  \begin{tabular}{lcccc}
    \hline
HJD-2\,450\,000. & $M1$ & mirror & 2-Gaussian & correlation \\
    \hline
9185.837 & -9.4 & -13.8 & -12.7 &  0.1 \\ 
9205.842 & -11.4 & -15.3 & -15.4 & -4.4 \\ 
9242.753 & -12.9 & -18.1 & -15.7 & -5.2 \\ 
9257.801 & -7.9 & -5.3 & -7.5 &  2.5 \\ 
9274.662 & -17.1 & -21.4 & -22.1 & -16.3 \\ 
\hline
  \end{tabular}

{\scriptsize The line was integrated from -250 to 250\,\kms\ for the moment method. Wings with normalized fluxes between 0.85 and 0.9 were used for the mirror method. The double Gaussians had center velocities $\pm a = \pm$150\,\kms.}
  \end{center}
\end{table}

\begin{table}
  \caption{Same as Table \ref{rv119682} but for V767\,Cen. \label{rv767}}
\begin{center}
  \begin{tabular}{lcccc}
    \hline
HJD-2\,450\,000. & $M1$ & mirror & 2-Gaussian & correlation \\
    \hline
9111.496 & -12.4 & -19.1 & -22.0 & -12.2 \\ 
9191.842 & -8.0 & -10.0 & -9.3 & -0.5 \\ 
9206.834 & -7.6 & -14.0 & -14.7 & -4.4 \\ 
9231.859 & -7.8 & -13.0 & -15.2 & -4.0 \\ 
9246.777 & -8.0 & -12.1 & -9.6 &  0.0 \\ 
\hline
  \end{tabular}

{\scriptsize The line was integrated from -200 to 200\,\kms\ for the moment method. Wings with normalized fluxes between 0.75 and 0.85 were used for the mirror method. The double Gaussians had center velocities $\pm a = \pm$50\,\kms.}
  \end{center}
\end{table}

\begin{table}
  \caption{Same as Table \ref{rv90563} but for CQ\,Cir. \label{rvcq}}
\begin{center}
  \begin{tabular}{lcccc}
    \hline
HJD-2\,450\,000. & $M1$ & mirror & 2-Gaussian & correlation \\
    \hline
9115.506 & -34.8 & -35.4 & -36.4 &  0.0 \\ 
9212.831 & -26.8 & -30.1 & -29.2 &  3.1 \\ 
9240.847 & -44.5 & -40.1 & -43.8 & -5.6 \\ 
9255.788 & -32.2 & -33.5 & -33.0 &  0.7 \\ 
9271.693 & -35.5 & -34.7 & -36.2 & -0.2 \\ 
\hline
  \end{tabular}

{\scriptsize The line was integrated from -300 to 300\,\kms\ for the moment method. Wings with normalized fluxes between 1.15 and 1.4 were used for the mirror method. The double Gaussians had center velocities $\pm a = \pm$200\,\kms.}
  \end{center}
\end{table}

\begin{table}
  \caption{Same as Table \ref{rv90563} but for HD\,157832. \label{rv157}}
\begin{center}
  \begin{tabular}{lcccc}
    \hline
HJD-2\,450\,000. & $M1$ & mirror & 2-Gaussian & correlation \\
    \hline
9108.595 & 16.4 & 17.8 & 18.1 &  0.0 \\ 
9123.548 & 16.8 & 15.5 & 16.5 & -0.8 \\ 
9138.537 & 11.1 &  9.1 &  8.9 & -8.4 \\ 
9156.510 &  7.0 &  7.7 &  7.5 & -10.3 \\ 
9246.856 &  7.0 &  7.9 &  7.6 & -11.3 \\ 
\hline
  \end{tabular}

{\scriptsize The line was integrated from -250 to 250\,\kms\ for the moment method. Wings with normalized fluxes between 1.1 and 1.3 were used for the mirror method. The double Gaussians had center velocities $\pm a = \pm$175\,\kms.}
  \end{center}
\end{table}

\begin{table}
  \caption{Same as Table \ref{rv90563} but for HD\,161103. \label{rv161}}
\begin{center}
  \begin{tabular}{lcccc}
    \hline
HJD-2\,450\,000. & $M1$ & mirror & 2-Gaussian & correlation \\
    \hline
9108.618 & -12.2 & -12.4 & -1.8 & -4.3 \\ 
9123.559 & -13.5 & -10.1 & -1.4 & -6.7 \\ 
9138.546 & -12.3 & -7.7 & -1.4 & -6.7 \\ 
9257.848 & -7.5 & -7.5 &  4.0 &  2.3 \\ 
9271.856 & -9.4 & -9.4 &  3.4 &  0.0 \\ 
\hline
  \end{tabular}

{\scriptsize The line was integrated from -200 to 200\,\kms\ for the moment method. Wings with normalized fluxes between 1.3 and 1.6 were used for the mirror method. The double Gaussians had center velocities $\pm a = \pm$150\,\kms.}
  \end{center}
\end{table}

\begin{table}
  \caption{Same as Table \ref{rv90563} but for V771\,Sgr. \label{rv771}}
\begin{center}
  \begin{tabular}{lcccc}
    \hline
HJD-2\,450\,000. & $M1$ & mirror & 2-Gaussian & correlation \\
    \hline
9109.590 & -11.6 & -11.9 & -12.0 &  0.0 \\ 
9123.602 & -13.3 & -14.0 & -13.9 & -2.1 \\ 
9138.574 & -13.7 & -14.2 & -14.4 & -3.0 \\ 
9258.870 & -7.8 & -7.4 & -7.4 &  4.2 \\ 
9282.837 & -6.0 & -5.9 & -5.9 &  6.2 \\ 
\hline
  \end{tabular}

{\scriptsize The line was integrated from -250 to 250\,\kms\ for the moment method. Wings with normalized fluxes between 1.2 and 1.6 were used for the mirror method. The double Gaussians had center velocities $\pm a = \pm$150\,\kms.}
  \end{center}
\end{table}

\begin{table}
  \caption{Same as Table \ref{rv90563} but for HD\,316568. \label{rv316}}
\begin{center}
  \begin{tabular}{lcccc}
    \hline
HJD-2\,450\,000. & $M1$ & mirror & 2-Gaussian & correlation \\
    \hline
9109.607 & -1.9 & -1.6 & -1.6 &  0.0 \\ 
9124.545 &  7.0 &  7.7 &  7.8 &  9.7 \\ 
9140.550 & 10.0 & 13.3 & 12.7 & 16.7 \\ 
9264.834 & 25.3 & 27.4 & 26.7 & 26.9 \\ 
9291.848 & 18.4 & 22.0 & 22.8 & 24.2 \\ 
\hline
  \end{tabular}

{\scriptsize The line was integrated from -200 to 250\,\kms\ for the moment method. Wings with normalized fluxes between 1.1 and 1.35 were used for the mirror method. The double Gaussians had center velocities $\pm a = \pm$150\,\kms.}
  \end{center}
\end{table}

\begin{table}
  \caption{Same as Table \ref{rv782a} but for V558\,Lyr. \label{rv558}}
\begin{center}
  \begin{tabular}{lcccc}
    \hline
HJD-2\,450\,000. & $M1$ & mirror & 2-Gaussian & correlation \\
    \hline
8941.979 & -36.4 & -52.2 & -52.7 &  0.0 \\ 
8960.952 & -38.1 & -48.7 & -50.1 &  1.6 \\ 
8982.971 & -30.1 & -41.9 & -42.1 & 11.6 \\ 
8996.964 & -29.1 & -40.2 & -41.2 & 10.3 \\ 
9019.846 & -39.8 & -52.3 & -54.1 &  1.2 \\ 
9024.855 & -38.1 & -55.2 & -56.6 & -1.4 \\ 
9045.757 & -40.7 & -56.2 & -56.6 & -4.1 \\ 
9082.668 & -30.2 & -41.6 & -43.3 & 10.0 \\ 
9116.590 & -39.8 & -55.9 & -57.7 & -4.7 \\ 
9137.597 & -37.3 & -51.9 & -52.0 & -2.3 \\ 
9156.558 & -26.1 & -38.0 & -40.1 & 11.3 \\ 
9175.565 & -30.4 & -42.9 & -44.6 &  7.4 \\ 
9306.920 & -28.9 & -43.6 & -45.3 &  3.2 \\ 
9325.934 & -21.1 & -31.0 & -32.2 & 12.6 \\ 
9436.745 & -19.1 & -28.1 & -27.2 & 12.2 \\ 
9495.598 & -19.2 & -21.8 & -18.9 &  5.2 \\ 
\hline
  \end{tabular}

{\scriptsize The line was integrated from -400 to 400\,\kms\ for the moment method. Wings with normalized fluxes between 1.5 and 2.5 were used for the mirror method. The double Gaussians had center velocities $\pm a = \pm$240\,\kms.}
  \end{center}
\end{table}

\begin{table}
  \caption{Same as Table \ref{rv782a} but for SAO\,49725. \label{rvsao}}
\begin{center}
  \begin{tabular}{lcccc}
    \hline
HJD-2\,450\,000. & $M1$ & mirror & 2-Gaussian & correlation \\
    \hline
\multicolumn{5}{l}{\it TIGRE}\\
9498.695 & -9.2 & -10.9 & -11.1 &  0.0 \\ 
\hline
\multicolumn{5}{l}{\it CARMENES}\\
8696.596 & -9.5 & -11.4 & -11.4 &  0.0 \\ 
8712.467 & -11.2 & -13.7 & -13.6 & -3.5 \\ 
8733.416 & -13.6 & -16.6 & -16.7 & -6.5 \\ 
8756.447 & -13.0 & -16.3 & -16.3 & -5.9 \\ 
8775.292 & -10.8 & -13.4 & -13.4 & -4.7 \\ 
8792.297 & -10.2 & -12.2 & -12.2 & -0.8 \\ 
9416.612 & -11.0 & -13.5 & -13.5 & -4.6 \\ 
9441.563 & -12.3 & -15.0 & -15.0 & -4.2 \\ 
9457.588 & -11.2 & -14.1 & -14.1 & -3.9 \\ 
9477.442 & -8.9 & -11.1 & -11.1 & -1.9 \\ 
9502.402 & -8.6 & -10.6 & -10.5 &  1.4 \\ 
\hline
  \end{tabular}

{\scriptsize The line was integrated from -300 to 300\,\kms\ for the moment method. Wings with normalized fluxes between 2.2 and 4.6 were used for the mirror method. The double Gaussians had center velocities $\pm a = \pm$120\,\kms.}
  \end{center}
\end{table}

\begin{table}
  \caption{Same as Table \ref{rv782a} but for V2156\,Cyg. \label{rv2156}}
\begin{center}
  \begin{tabular}{lcccc}
    \hline
HJD-2\,450\,000. & $M1$ & mirror & 2-Gaussian & correlation \\
    \hline
\multicolumn{5}{l}{\it TIGRE}\\
9336.908 & -8.1 & -10.6 & -10.3 &  0.0 \\ 
9437.820 & -12.9 & -15.3 & -15.4 & -5.0 \\ 
9497.625 & -10.1 & -12.1 & -12.2 & -1.7 \\ 
\hline
\multicolumn{5}{l}{\it CARMENES}\\
8696.617 & -7.0 & -8.8 & -7.2 &  0.0 \\ 
8712.486 & -5.2 & -6.0 & -5.1 &  0.7 \\ 
8756.464 & -11.1 & -14.3 & -13.6 & -4.7 \\ 
8775.313 & -13.8 & -16.8 & -16.1 & -8.2 \\ 
8792.342 & -15.8 & -19.0 & -18.2 & -11.0 \\ 
9416.637 & -17.2 & -19.9 & -20.0 & -13.9 \\ 
9438.586 & -12.1 & -14.6 & -14.6 & -7.9 \\ 
9457.614 & -9.2 & -9.4 & -9.0 & -5.7 \\ 
9477.461 & -6.4 & -6.8 & -6.6 & -2.8 \\ 
9502.428 & -12.3 & -13.7 & -13.3 & -7.3 \\ 
\hline
  \end{tabular}

{\scriptsize The line was integrated from -475 to 475\,\kms\ for the moment method. Wings with normalized fluxes between 2.2 and 3.6 were used for the mirror method. The double Gaussians had center velocities $\pm a = \pm$175\,\kms.}
  \end{center}
\end{table}

\begin{table}
  \caption{Same as Table \ref{rv782a} but for V810\,Cas. \label{rv810}}
\begin{center}
  \begin{tabular}{lcccc}
    \hline
HJD-2\,450\,000. & $M1$ & mirror & 2-Gaussian & correlation \\
    \hline
\multicolumn{5}{l}{\it TIGRE}\\
9418.885 & -31.8 & -36.8 & -36.1 &  0.0 \\ 
9446.758 & -20.6 & -23.2 & -23.4 & 15.0 \\ 
9496.744 & -29.9 & -35.2 & -34.4 &  3.8 \\ 
\hline
\multicolumn{5}{l}{\it CARMENES}\\
8696.635 & -18.3 & -20.9 & -21.2 &  0.0 \\ 
8712.641 & -21.4 & -24.8 & -24.7 & -2.4 \\ 
8733.664 & -29.5 & -34.1 & -33.6 & -13.4 \\ 
8756.478 & -21.7 & -27.5 & -27.6 & -2.2 \\ 
8775.335 & -18.3 & -20.6 & -20.7 & -0.2 \\ 
9416.652 & -30.6 & -35.8 & -35.4 & -15.0 \\ 
9438.557 & -22.3 & -27.5 & -27.3 & -2.2 \\ 
9457.632 & -18.8 & -21.0 & -21.5 & -0.3 \\ 
9477.476 & -26.3 & -30.5 & -31.0 & -10.0 \\ 
9502.448 & -25.6 & -31.8 & -31.8 & -7.3 \\ 
\hline
  \end{tabular}

{\scriptsize The line was integrated from -500 to 500\,\kms\ for the moment method. Wings with normalized fluxes between 1.6 and 2.8 were used for the mirror method. The double Gaussians had center velocities $\pm a = \pm$220\,\kms.}
  \end{center}
\end{table}

\bsp	
\label{lastpage}
\end{document}